\documentclass[10pt,twocolumn,letterpaper]{article}

\usepackage[pagenumbers]{iccv} % To force page numbers, e.g. for an arXiv version
\usepackage{algorithm, algorithmicx, algpseudocode}
\usepackage[accsupp]{axessibility}  % Improves PDF readability for those with disabilities.

\newcommand\awesomemesh{\textcolor{black}{\sc\textbf{V2M4}}\xspace}

\definecolor{iccvblue}{rgb}{0.21,0.49,0.74}
\usepackage[pagebackref,breaklinks,colorlinks,allcolors=iccvblue]{hyperref}
\usepackage{multirow}

\def\paperID{****} % *** Enter the Paper ID here
\def\confName{ICCV}
\def\confYear{2025}

\title{V2M4: 4D Mesh Animation Reconstruction from a Single Monocular Video}

\author{Jianqi Chen~~~~~Biao Zhang~~~~~Xiangjun Tang~~~~~Peter Wonka\\
KAUST\\
{\tt\small \{jianqi.chen, biao.zhang, xiangjun.tang, peter.wonka\}@kaust.edu.sa}
}

\begin{document}
% \maketitle
\twocolumn[{%
    \renewcommand\twocolumn[1][]{#1}%
    \maketitle
    \begin{center}
        \centering
        \captionsetup{type=figure}
        \includegraphics[width=\textwidth]{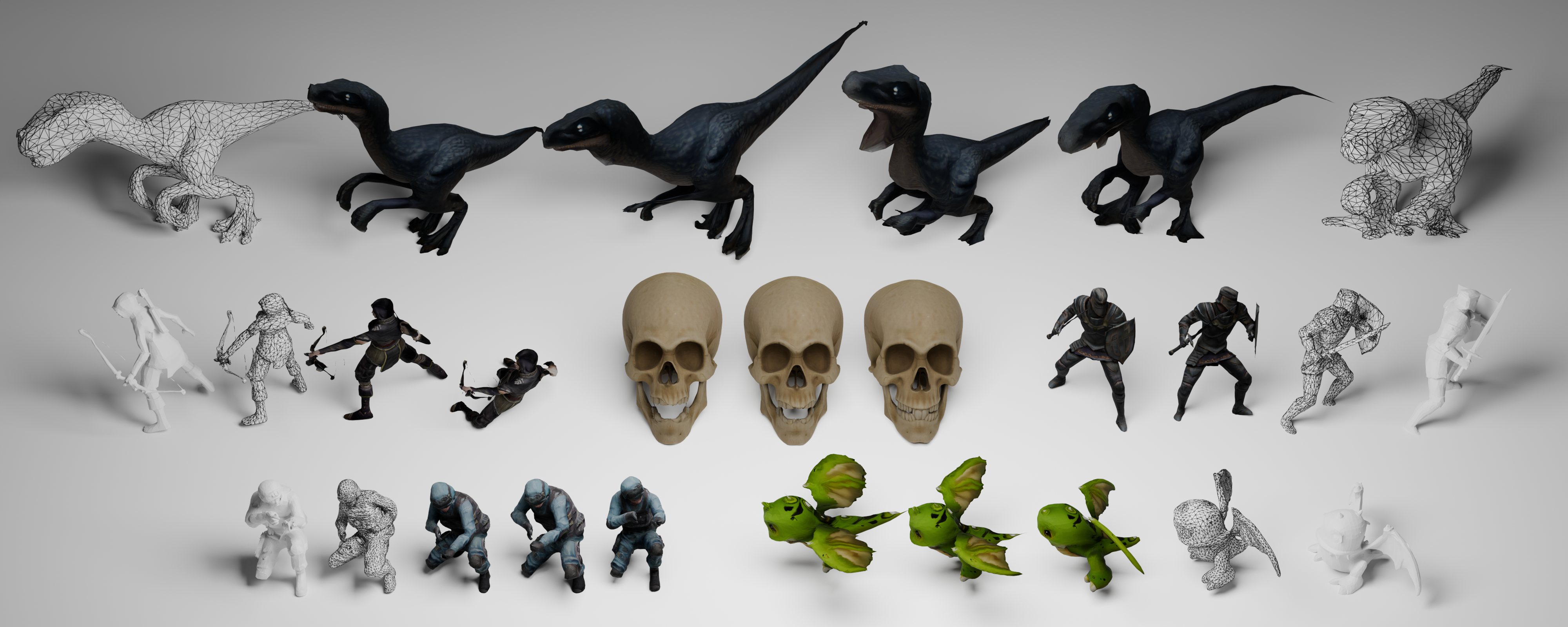}
        \vspace{-20pt}
        \captionof{figure}{Given a monocular video, our method can output a high-quality 4D mesh asset readily usable in graphics and game software.}
        \label{fig:teaser}
    \end{center}%
}]

% =============================================== ABSTRACT ==================================================
\begin{abstract}
We present \awesomemesh, a novel 4D reconstruction method that directly generates a usable 4D mesh animation asset from a single monocular video. Unlike existing approaches that rely on priors from multi-view image and video generation models, our method is based on native 3D mesh generation models. Naively applying 3D mesh generation models to generate a mesh for each frame in a 4D task can lead to issues such as incorrect mesh poses, misalignment of mesh appearance, and inconsistencies in mesh geometry and texture maps. To address these problems, we propose a structured workflow that includes camera search and mesh reposing, condition embedding optimization for mesh appearance refinement, pairwise mesh registration for topology consistency, and global texture map optimization for texture consistency. Our method outputs high-quality 4D animated assets that are compatible with mainstream graphics and game software. Experimental results across a variety of animation types and motion amplitudes demonstrate the generalization and effectiveness of our method. \url{https://windvchen.github.io/V2M4}.
\end{abstract}

% =============================================== Introduction ==================================================
\vspace{-20pt}
\section{Introduction}
\label{sec:Introduction}

Creating 4D mesh animations is a challenging task. Unlike the synthesis of 3D geometry or 2D images, the scarcity of 4D mesh animations makes it difficult to employ a learning-based framework. Monocular videos provide rich motion cues, making them a potential source for animation reconstruction. However, reconstructing 4D mesh animations from single-view videos remains an underexplored problem, posing challenges in addressing the ambiguity in single-view video and maintaining both visual and topological consistency over time.

Existing methods~\cite{ren2025l4gm, zeng2024stag4d, jiang2023consistent4d, xie2024sv4d, zhang20254diffusion, zhao2023animate124, jiang2024animate3d, sun2024eg4d, yang2024diffusion, ren2023dreamgaussian4d, wu2024sc4d} have explored animation reconstruction based on implicit representations such as NeRF~\cite{mildenhall2021nerf} or topology-independent representations like Gaussian Splatting~\cite{kerbl20233d}. To mitigate occlusions and ambiguities in single-view videos, these approaches leverage knowledge embedded in multi-view image diffusion~\cite{voleti2024sv3d, liu2023zero, shi2023mvdream} and video diffusion~\cite{ho2022video, blattmann2023stable}. While these methods are viable alternatives, they face disadvantages, such as the difficulty of ensuring 3D consistency in multi-view image diffusion and video diffusion. Additionally, converting these implicit and topology-independent representations into a mesh representation remains challenging. These factors complicate the use of these methods to obtain a mesh animation that maintains a fine-grained geometric shape and preserves inter-frame topological consistency.

In this work, we propose \textit{\awesomemesh}, a method that directly reconstructs a 4D animated mesh from a fixed-viewpoint monocular video. Unlike existing approaches that rely on multi-view image and video diffusion models, our method leverages recent advancements in native 3D mesh reconstruction ~\cite{xiang2024structured, huang2025spar3d}, enabling explicit mesh generation with good shape without the need to convert from other representations. A naive approach to using 3D mesh reconstruction for 4D animation involves generating a mesh for each frame independently. However, this strategy presents several challenges. Chief among these is the uncertainty of the generated mesh pose, where random face orientations and positions lead to misalignment with the video frames. Additionally, there are issues of appearance misalignment, where the generated meshes fail to match the object's appearance in the video. Furthermore, inconsistencies in geometry and topology result in varying mesh structures across frames, and differences in texture maps disrupt the visual continuity of the animation.

To tackle these challenges, we introduce a structured approach that effectively extends the 3D mesh reconstruction model to the 4D mesh animation task. \textbf{First}, we develop a mesh reposing strategy. In this step, we search for camera poses for each reconstructed mesh to ensure semantic alignment with the input video frame by combining particle swarm optimization with a dense stereo reconstruction model's prior. We then re-pose the meshes by applying inverse camera transformations. \textbf{Second}, we refine the appearance of the meshes by optimizing the negative condition embedding of the 3D reconstruction model, enhancing consistency with the video frames. \textbf{Third}, we enforce geometry consistency by selecting the first-frame mesh as a \textit{rest pose} and introducing a global-to-local mesh registration approach using differential rendering and geometric constraints. \textbf{Fourth}, we optimize a global texture map to ensure texture consistency based on the rendering views of the reconstructed meshes. \textbf{Finally}, we keyframe the meshes and interpolate vertex positions between frames to achieve smoother animation, integrating the results into a GLTF~\cite{glTF} animated file, which is compatible with mainstream graphics software. Fig.~\ref{fig:teaser} displays some of our results. Our main contributions are as follows:

\begin{itemize}

\item We introduce the use of native 3D mesh reconstruction models for 4D animated asset generation, enabling the direct creation of explicit 4D meshes with consistent geometry, shared textures, and video-aligned object motions from a fixed-viewpoint monocular video.

\item  We propose \textit{\awesomemesh}, a practical and efficient framework that addresses key challenges associated with using 3D mesh reconstruction models for 4D mesh reconstruction, including incorrect mesh positioning, appearance misalignment, geometry inconsistencies, and texture inconsistencies.

\item We conduct extensive evaluations on diverse animations, including humanoid and animal motions, covering a range of motion amplitudes from subtle movements to large-scale motions. The results demonstrate the effectiveness and generalizability of our approach.

\end{itemize}

% =============================================== Related Works ==================================================
\section{Related Works}
\label{sec:Related Works}

\paragraph{Multi-view and Video Diffusion Models.} Building on the success of image diffusion models~\cite{ho2020denoising, rombach2022high}, recent works have increasingly focused on the generation of other modalities, such as multi-view images~\cite{shi2023mvdream, liu2023zero, voleti2024sv3d} and videos~\cite{singer2022make, ho2022video, guo2023animatediff}. In multi-view diffusion models, networks predict the 3D structure of an object from an input image. For instance, Zero123~\cite{liu2023zero} can generate a rendering of an object from a novel camera viewpoint by taking an input image and a target camera pose transition. Subsequent works, such as ImageDream~\cite{wang2023imagedream} and SyncDream~\cite{liu2023syncdreamer}, further advance this direction by generating multi-view images in a single forward pass from a single input image. Video diffusion models have seen even more rapid development, driven by the availability of large-scale video datasets. Methods such as Dynamicrafter~\cite{xing2024dynamicrafter} and Stable Video Diffusion~\cite{blattmann2023stable} can generate high-quality videos and enable smooth interpolation between frames.

\begin{table}[t]
\resizebox{\columnwidth}{!}{\begin{tabular}{@{}lccc@{}}
\toprule
\textbf{Method} &
  \textbf{3D Prior Source} &
  \textbf{\begin{tabular}[c]{@{}c@{}}Reconstruction\\ Process\end{tabular}} &
  \textbf{\begin{tabular}[c]{@{}c@{}}Output\\ Format\end{tabular}} \\ \midrule
\rowcolor[HTML]{EFEFEF} 
Consistent4D    & \cellcolor[HTML]{EFEFEF}MV Image and Video Diffusion & Optimization                     & NeRF            \\
STAG4D      & MV Image Diffusion                   & Optimization   & GS         \\
\rowcolor[HTML]{EFEFEF} 
Diffsuion$^2$   & \cellcolor[HTML]{EFEFEF}MV Image and Video Diffusion & Optimization                   & GS              \\
EG4D        & MV Image and Video Diffusion         & Optimization & GS         \\
\rowcolor[HTML]{EFEFEF} 
Animate3D   & MV-Video Diffusion                   & Optimization   & GS         \\
4Diffusion  & MV-Video Diffusion                   & Optimization   & NeRF       \\
\rowcolor[HTML]{EFEFEF} 
Diffusion4D & MV-Video Diffusion                   & Optimization & GS         \\
SV4D        & MV-Video Diffusion                   & Optimization & NeRF       \\
\rowcolor[HTML]{EFEFEF} 
AR4D        & MV Image Diffusion & Recon Network     & GS         \\
L4GM        & MV Image Diffusion                      & Recon Network     & GS         \\
\rowcolor[HTML]{EFEFEF} 
DreamMesh4D & MV Image Diffusion                   & Optimization   & Mesh \& GS \\ \midrule
\textbf{\awesomemesh} (Ours)   & \textbf{Native 3D Mesh Diffuion Model}         & \textbf{-}              & \textbf{Mesh}   \\ \bottomrule
\end{tabular}}
\caption{\textbf{Comparison of \awesomemesh with Existing 4D Reconstruction Methods.} ``-" indicates the absence of an intermediate reconstruction phase. Not all existing 4D methods are listed, but other approaches are generally similar to those included in the table.}
\label{Tab:differences}
\vspace{-10pt}
\end{table}

\vspace{-10pt}
\paragraph{Video-to-4D Reconstruction Models.} Due to the scarcity of 4D animated assets, existing research primarily leverages priors learned from multi-view and video generative models to ensure spatial and temporal consistency in reconstructed 4D animations. Consistent4D~\cite{jiang2023consistent4d} employs Score Distillation Sampling (SDS) with signals from both pretrained multi-view and video diffusion models. STAG4D~\cite{zeng2024stag4d} follows a similar approach but introduces an inter-frame key-value mixing mechanism during the denoising process to enhance sequence consistency. To avoid SDS, AR4D~\cite{zhu2025ar4d} instead proposes deforming the initial frame's Gaussian Splats while utilizing priors from multi-view generation models. Diffusion$^2$~\cite{yang2024diffusion} and EG4D~\cite{sun2024eg4d} modify the sampling or attention mechanisms to integrate prior knowledge from multi-view image and video diffusion models without relying on SDS. Animate3D~\cite{jiang2024animate3d} and 4Diffusion~\cite{zhang20254diffusion} train multi-view video diffusion models that generate more consistent images for improved SDS optimization. Similarly, Diffusion4D~\cite{liang2024diffusion4d} and SV4D~\cite{xie2024sv4d} also train multi-view video generation models but directly optimize Gaussian Splatting or NeRF on the generated images without SDS. L4GM~\cite{ren2025l4gm} takes a different approach by training a network that outputs 4D Gaussian Splats given multiview images. However, all of these methods do not output meshes, as discussed in Sec.~\ref{sec:Introduction}. An alternative approach, DreamMesh4D~\cite{li2025dreammesh4d}, introduces a hybrid Gaussian-Mesh representation capable of outputting meshes, though it still relies on multi-view and video diffusion model priors and adopts SDS optimization. Other works~\cite{yang2022banmo, yang2023reconstructing, song2023total} also attempt to reconstruct meshes from videos but require significantly more input, such as multiple video clips, rendering camera poses, skeletons, or categorical dense feature descriptors, rather than just a single monocular video.

\vspace{-10pt}
\paragraph{Large Native 3D Mesh Generation Models.} Unlike previous indirect 3D generative methods, which either generate multi-view images~\cite{wang2023imagedream, liu2023syncdreamer,tang2024lgm} or implicit representations ~\cite{yan2022shapeformer,Biao_2022_3DILG, Biao_2023_VecSet,zheng2023locally}, recent work ~\cite{xiang2024structured, huang2025spar3d} focuses on directly generating usable meshes by adopting differentiable meshing techniques such as FlexiCubes~\cite{shen2023flexicubes} or DMTet~\cite{shen2021dmtet}. Instead, they directly predict grid attributes and generate explicit meshes. By training on large collections of synthesized 3D assets, the method achieves impressive generation results conditioned on images.

\vspace{-10pt}
\paragraph{Ours versus Others.} Different from most existing methods that rely on priors from multi-view image or video diffusion models and use NeRF or Gaussian Splatting as representations, we pioneer the use of native 3D mesh generation models for 4D animation reconstruction. Our approach only requires a single monocular video with a fixed camera pose as input, without the need for SDS, and outputs a 4D mesh animation file with consistent geometry and texture. Table~\ref{Tab:differences} illustrates how our approach differs from previous works.

\begin{figure*}[!t]
  \centering
   \includegraphics[width=\linewidth]{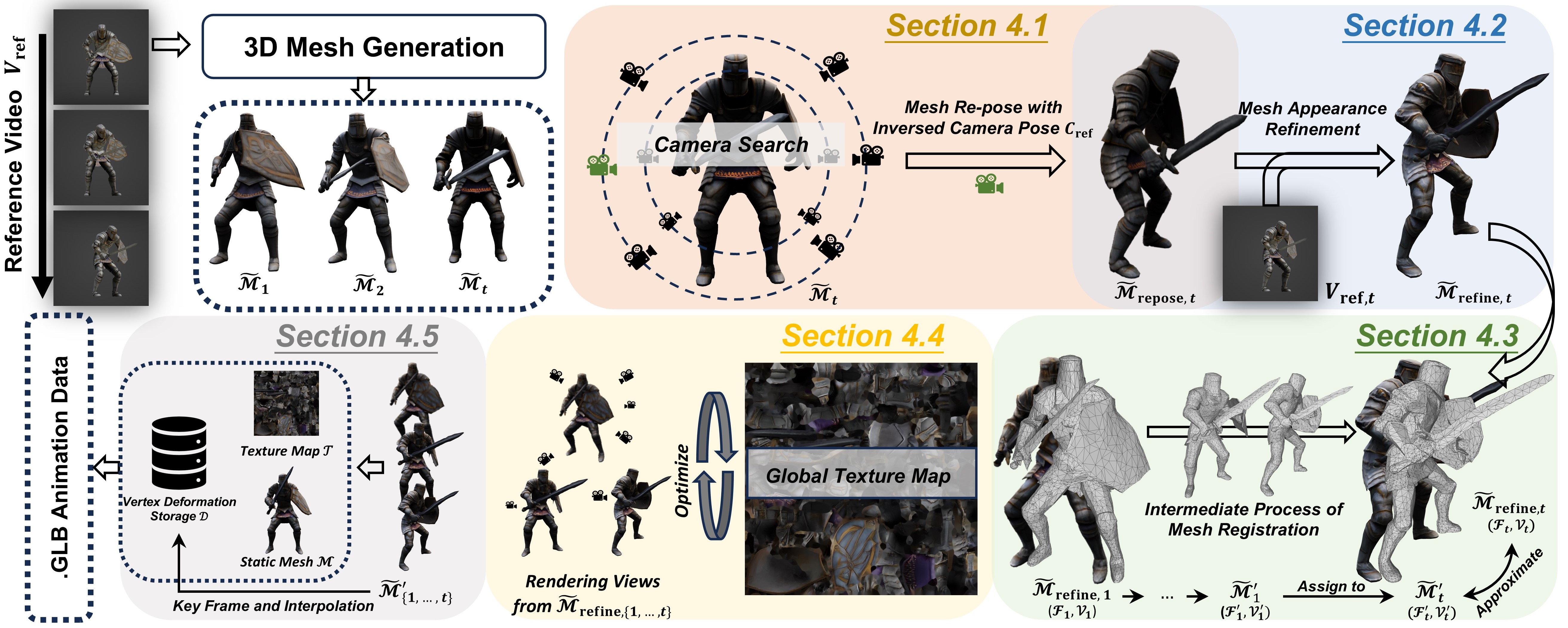}
   \vspace{-20pt}
   \caption{\textbf{The Workflow of \textit{\awesomemesh}.} Upon receiving a video sequence \( V_{\text{ref}} \), we generate coarse 3D meshes for each frame, denoted as \(\widetilde{\mathcal{M}}_{\{1, \dots, t\}}\). These initial meshes do not accurately capture the object's movement and appearance as depicted in the input video, and they exhibit inconsistencies in topology and texture. Our method employs a five-stage process: (1) Repose the mesh to accurately reflect object movement; (2) Refine the mesh appearance based on the reference video frames; (3) Ensure inter-mesh topology consistency through mesh registration; (4) Optimize a globally shared texture map across all meshes; (5) Keyframe the meshes, perform interpolation, and convert them into a directly usable 4D animation asset.}
   \label{fig:framework}
   \vspace{-10pt}
\end{figure*}

% =============================================== Preliminaries ==================================================
\section{Preliminaries}
\label{sec:Preliminaries}

Our method builds upon TRELLIS~\cite{xiang2024structured}, a recently introduced and powerful 3D generation model (we also evaluate our method using more advanced 3D generators, as detailed in Appendix~\ref{appendix:Ablation Studies}). TRELLIS encodes 3D assets into a structured latent representation (SLAT), which consists of a sparse 3D grid embedding both geometric and appearance information. The 3D reconstruction process in TRELLIS can be broadly divided into two phases. In the first phase, TRELLIS employs rectified flow ~\cite{esser2024scaling} to generate a 3D voxel grid, where the voxel values indicate the object's occupancies. By applying a threshold, TRELLIS extracts a sparse voxel representation. In the second phase, a sparse rectified flow transformer is used to generate the object's SLAT from this sparse voxel grid. Both phases are conditioned on visual features extracted from the input image using DINOv2 ~\cite{oquab2023dinov2}, along with a negative condition based on the classifier-free technique ~\cite{ho2022classifier}. The SLAT can then be processed by three independent decoders, which generate meshes, Gaussian Splats ~\cite{kerbl20233d}, and Radiance Fields ~\cite{gao2023strivec}, respectively. For mesh decoding, TRELLIS predicts both the signed distance field (SDF) grid and the parameters of differentiable FlexiCubes ~\cite{shen2023flexicubes}, enabling the conversion from the SDF into a mesh. Instead of using the predicted mesh vertex colors from the mesh decoder, TRELLIS employs xatlas ~\cite{xatlas} to compute UV maps of the mesh and then bakes a high-resolution texture map using multi-view observations of its Gaussian Splats.

% =============================================== Method ==================================================
\section{Method}
\label{sec:Method}

Given a reference monocular video sequence of length \( T \), denoted as \( V_{\text{ref}} = \{V_{\text{ref},t}\}_{t=1}^T \), which captures an object from a fixed camera pose, our method generates a directly usable 4D animated asset. This asset comprises a static mesh \(\mathcal{M}\) with vertices \(\mathcal{V} = \{\mathbf{v}_1, \mathbf{v}_2, \cdots, \mathbf{v}_N\}\) (or simply $\mathcal{V}\in\mathbb{R}^{N\times 3}$) and faces \(\mathcal{F} \subset \mathcal{V}\times \mathcal{V}\times \mathcal{V}\), a texture map \(\mathcal{T}\), and a deformation tensor \(\mathcal{D} \in \mathbb{R}^{T \times N \times 3}\) that records the vertices' deformations over time. Initially, each video frame is processed through an image-conditioned 3D generation model, resulting in a set of initial 3D meshes, denoted as \(\widetilde{\mathcal{M}}_{\text{init}} = \{\widetilde{\mathcal{M}}_1, \widetilde{\mathcal{M}}_2, \dots, \widetilde{\mathcal{M}}_T\}\). These meshes exhibit inconsistencies with \( V_{\text{ref}} \) as well as inter-mesh texture and geometry discrepancies, rendering them unsuitable for direct conversion into a usable 4D asset. To address these issues, our method employs a five-step workflow: repositioning the mesh for accurate object movement (Sec.~\ref{subsec:Semantic Camera Alignment}), refining object appearance (Sec.~\ref{subsec:Mesh Appearance Alignment}), ensuring inter-frame mesh geometry consistency (Sec.~\ref{subsec:Consistent_Geometry}), maintaining inter-frame mesh texture consistency (Sec.~\ref{subsec:Consistent_Geometry}), and ultimately converting the 3D assets into 4D assets (Sec.~\ref{subsec:Mesh Interpolation and 4D Asset Conversion}). The complete workflow is illustrated in Fig.~\ref{fig:framework}.

\subsection{Camera Search and Mesh Re-Pose}
\label{subsec:Semantic Camera Alignment}

Due to the setup of training TRELLIS, the initial meshes \(\widetilde{\mathcal{M}}_{\text{init}}\) often tend to have a canonical face orientation along one axis (as illustrated in Fig.~\ref{fig:framework}), and their positions may also sometimes be canonicalized to the origin of the coordinate system. This results in a significant loss of object motion information, such as translation and rotation. To recover these motions, we propose first determining the camera pose \(\mathcal{C} \in \mathbb{R}^6 = \{\mathrm{yaw}, \mathrm{pitch}, \mathrm{radius}, \mathrm{lookat}_{x,y,z}\}\) for each reconstructed mesh \(\widetilde{\mathcal{M}}_t\), ensuring the rendered view is semantically similar to the corresponding video frame. Then, we can invert the camera pose and apply the transformation to the mesh, effectively recovering mesh motion from the camera motion.

Since directly optimizing camera parameters via gradient descent from a random starting point using differentiable rendering tools like Nvdiffrast ~\cite{Laine2020diffrast} often leads to poor local minima, we propose a more robust camera pose search strategy. Specifically, we begin by sampling a large number of camera positions around the object and selecting the top \( n \) positions that exhibit high similarity to the reference video frame, denoted as \({V}_{\text{ref},t}\). We then utilize a pretrained dense stereo model to process the rendering views under these \( n \) cameras and \({V}_{\text{ref},t}\), obtaining the predicted point clouds for each view. Next, we design to extract the camera pose \(\mathcal{C}_{\text{ref},t,\text{DUSt3R}}\) for the reference image from these point clouds. This pose \(\mathcal{C}_{\text{ref},t,\text{DUSt3R}}\) is added back to the initial pool of camera pose candidates, and we employ the Particle Swarm Optimization (PSO) algorithm to refine this into a more accurate camera pose \(\mathcal{C}_{\text{ref},t,\text{PSO}}\). Subsequently, we further refine the camera pose using gradient descent optimization, starting from the improved point \(\mathcal{C}_{\text{ref},t,\text{PSO}}\), to obtain the final \(\mathcal{C}_{\text{ref},t}\). Finally, we derive the camera motion from \(\mathcal{C}_{\text{ref},t}\), invert it, and apply this transformation to \(\widetilde{\mathcal{M}}_t\) to obtain \(\widetilde{\mathcal{M}}_{\text{repose}, t}\). The entire workflow is detailed in Algorithm~\ref{alg:camera_alignment}.

Below, we provide more details about the three key components involved.

\begin{algorithm}[t]
\caption{Camera Search and Mesh Re-Pose}
\label{alg:camera_alignment}
\begin{algorithmic}[1]
    \Require Initial mesh \(\widetilde{\mathcal{M}}_t\), reference frame \(V_{\text{ref},t}\) at time \(t\), pretrained dense stereo reconstruction model DUSt3R
    
    \State \textbf{Initialize} a set of sampled camera poses \(P_0\)
    \State Select the top \(n\) camera poses based on similarity to \(V_{\text{ref},t}\). Render views under these poses and input them, along with \(V_{\text{ref},t}\), into the DUSt3R
    
    \State \textbf{DUSt3R Estimation:} Align DUSt3R-predicted point clouds with ground-truth point clouds
    \State Optimize the camera extrinsic matrix to derive \(\mathcal{C}_{\text{ref},t,\text{DUSt3R}}\)
    
    \State \textbf{PSO Search:} Add \(\mathcal{C}_{\text{ref},t,\text{DUSt3R}}\) to the top \(K\) camera poses from \(P_0\), then apply PSO
    \State Select the best camera pose \(\mathcal{C}_{\text{ref},t,\text{PSO}}\) after iterations
    
    \State \textbf{Gradient Descent Refinement:} Refine by minimizing mask discrepancy with \(V_{\text{ref},t}\)
    \State Obtain the final optimized camera pose \(\mathcal{C}_{\text{ref},t}\)
    
    \State \textbf{Mesh Re-pose:} Apply the inverse of \(\mathcal{C}_{\text{ref},t}\) to \(\widetilde{\mathcal{M}}_t\) to obtain \(\widetilde{\mathcal{M}}_{\text{repose}, t}\)
    
    \State \Return Re-posed mesh \(\widetilde{\mathcal{M}}_{\text{repose}, t}\)
\end{algorithmic}
\end{algorithm}

\vspace{-10pt}
\paragraph{Camera Search Using Particle Swarm Optimization.} We use PSO~\cite{kennedy1995particle} algorithm to search for the aligned camera pose. To effectively explore the search space and avoid the local minima problem encountered with direct gradient descent, we sample $P$ particles (camera poses) around the object. The sampling strategy is designed to ensure that the $\mathrm{yaw}$ and $\mathrm{pitch}$ values are uniformly distributed over a unit sphere, while the $\mathrm{radius}$ is sampled according to the surface area of the sphere along with its radius. The $\mathrm{lookat}_{x,y,z}$ values are uniformly sampled within a predefined range. For the objective function, we utilize DreamSim~\cite{fu2023dreamsim}, which is more effective at evaluating the similarity between the rendered view of the imperfectly reconstructed mesh and the reference video frame, compared to conventional methods like MSE or LPIPS~\cite{zhang2018unreasonable}. The positions of the particles (camera pose values) are iteratively updated based on both their individual best-known position and the best-known position of the entire swarm, with added noise randomness. The final best one is selected as the reference image’s camera pose $\mathcal{C}_{\text{ref},t,\text{PSO}}$. 

\begin{figure}[!t]
  \centering
   \includegraphics[width=\linewidth]{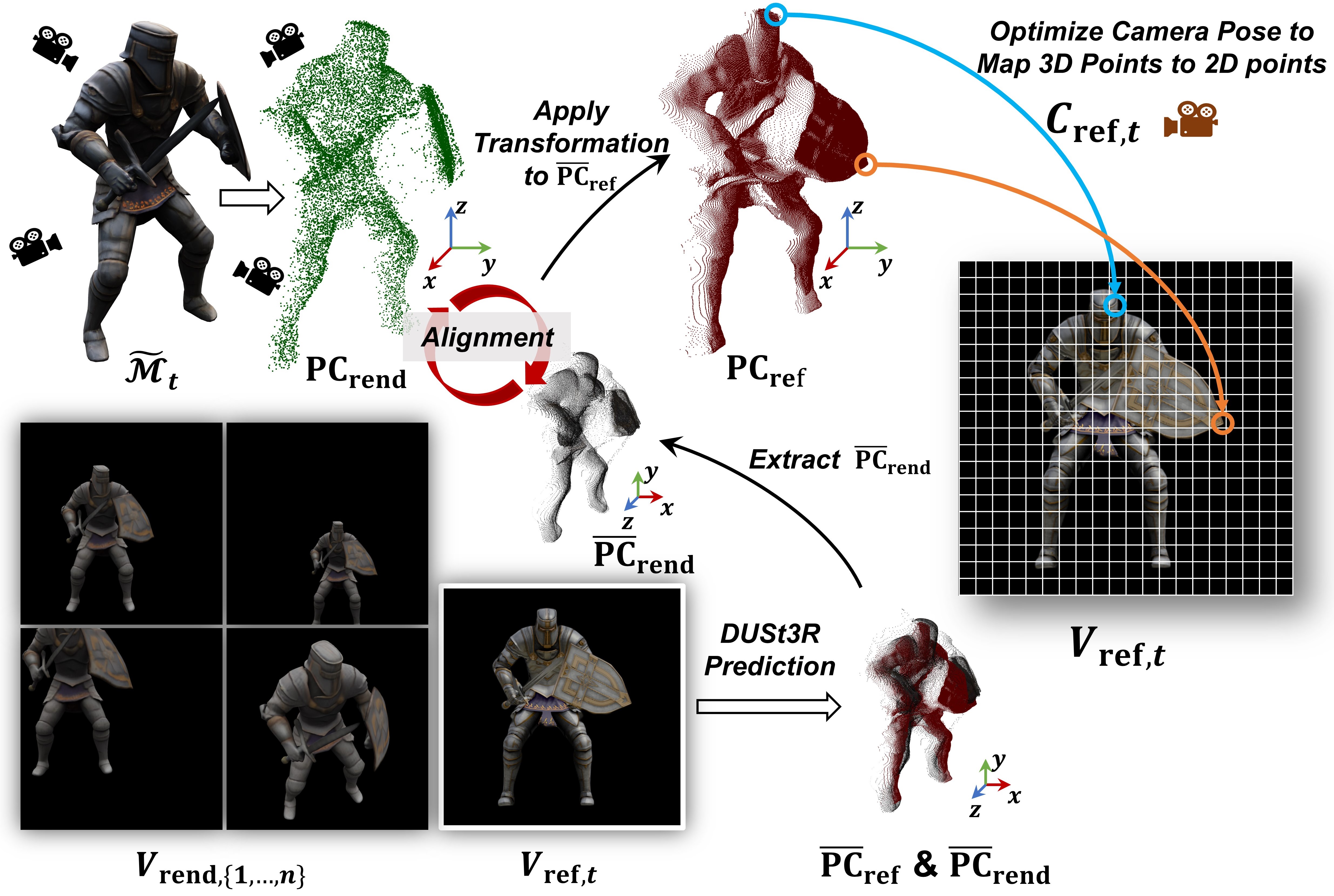}
   \vspace{-20pt}
   \caption{\textbf{Camera Pose Extraction from Dense Stereo Reconstruction for \( V_{\text{ref},t} \).} We first obtain estimated point clouds \(\overline{\mathrm{PC}}_{\text{ref}}\) and \(\overline{\mathrm{PC}}_{\text{rend}}\) from the DUSt3R model for reference video frame \( V_{\text{ref},t} \) and rendering views \( V_{\text{rend}} \). We then align \(\overline{\mathrm{PC}}_{\text{rend}}\) with the point clouds \({\mathrm{PC}}_{\text{rend}}\) obtained from mesh \(\widetilde{\mathcal{M}}_t\) during rasterization (Note the differences in axis direction and scale depicted in the figure.). This transformation is then applied to \(\overline{\mathrm{PC}}_{\text{ref}}\), resulting in the transformed \({\mathrm{PC}}_{\text{ref}}\). Finally, we optimize to determine the camera pose \(\mathcal{C}_{\text{ref},t}\) that ensures accurate mapping from 3D to 2D points.}
   \label{fig:dust3r}
   \vspace{-10pt}
\end{figure}

\vspace{-10pt}
\paragraph{Knowledge Utilization from Dense Stereo Reconstruction Model.} To enhance the robustness of the PSO search, we leverage prior knowledge from a pretrained dense stereo model, specifically DUSt3R~\cite{wang2024dust3r}. DUSt3R can predict inter-frame camera motion and 3D point positions corresponding to each view's 2D pixels. While a straightforward approach would involve feeding both the reference video frame \( V_{\text{ref},t} \) and a set of rendered views \( V_{\text{rend},\{1, \dots, n\}} \) around \(\widetilde{\mathcal{M}}_t\) into DUSt3R and using the predicted relative camera motion to estimate the camera pose of \( V_{\text{ref},t} \), this poses challenges. Differences in camera models and coordinate settings, such as scale and orientation, complicate matrix conversions, and semantic discrepancies between \( V_{\text{ref},t} \) and \( V_{\text{rend},\{1, \dots, n\}} \) can lead to inaccurate motion predictions. To overcome these issues, we focus on the point clouds estimated by DUSt3R instead of relying on the predicted camera motions. Let \(\overline{\mathrm{PC}}_{\text{ref}}\) and \(\overline{\mathrm{PC}}_{\text{rend},\{1, \dots, n\}}\) represent the predicted point clouds for \( V_{\text{ref},t} \) and \( V_{\text{rend},\{1, \dots, n\}} \), respectively. Given that the camera model, camera poses, and \(\widetilde{\mathcal{M}}_t\) are known for rendering \( V_{\text{rend},\{1, \dots, n\}} \), we can derive the ground-truth point clouds in our coordinate system, denoted as \(\mathrm{PC}_{\text{rend},\{1, \dots, n\}}\). We align the predicted point clouds with the ground truth using a global transformation (rotation, translation, and scale) optimized via Chamfer distance ~\cite{fan2017point}. Applying this transformation to \(\overline{\mathrm{PC}}_{\text{ref}}\), we obtain \({\mathrm{PC}}_{\text{ref}}\). Ideally, with the correct camera pose \(\mathcal{C}_{\text{ref},t}\), the 3D positions of \({\mathrm{PC}}_{\text{ref}}\) should project onto the 2D pixel coordinates of \( V_{\text{ref},t} \). We optimize the camera's extrinsic matrix to ensure this alignment, thereby deriving the reference image camera pose \(\mathcal{C}_{\text{ref},t}\) from the extrinsic matrix.

\vspace{-10pt}
\paragraph{Camera Refinement via Gradient Descent.} Given an approximately aligned camera pose where the rendering view matches \( V_{\text{ref},t} \), we further refine this alignment by applying gradient descent optimization on the camera pose. Specifically, we minimize the discrepancy between the mask region of the rendering view—defined as the pixels covered by the object during rasterization—and the object region in $V_{\text{ref},t}$. With a well-initialized starting point, the gradient descent efficiently converges to a near-optimal solution for $\mathcal{C}_{\text{ref},t}$.

\subsection{Mesh Appearance Refinement via Negative Condition Embedding Optimization} 
\label{subsec:Mesh Appearance Alignment}

With $\mathcal{C}_{\text{ref},t}$, the object’s viewpoint is now synchronized with the video frame. This alignment allows us to leverage supervision from $V_{\text{ref},t}$ to improve the consistency of the mesh appearance with the input video. Inspired by techniques in image editing ~\cite{mokady2023null, chen2025zero}, we propose optimizing the negative condition embedding during the second phase of TRELLIS inference (described in Sec.~\ref{sec:Preliminaries}). This optimization preserves the priors learned by the network, as opposed to directly optimizing the final output SLAT. We begin optimization after a few denoising steps with the sparse flow transformer, ensuring that the object’s structure is roughly generated. The optimization objective is to maximize similarity (measured by DreamSim, MSE, and LPIPS, with a regularization term to prevent significant deviation from the initial values) between the rendered view of the decoded SLAT (\textbf{both mesh and Gaussian splats}) under the camera pose $\mathcal{C}_{\text{ref},t}$ and the reference video frame $V_{\text{ref},t}$. The optimization is performed over multiple iterations for each denoising timestep. We denote the refined mesh as \(\widetilde{\mathcal{M}}_{\text{refine}, t}\).

\subsection{Consistent Topology via Iterative Pairwise Registration}    
\label{subsec:Consistent_Geometry}  

Due to the inherent randomness of the generation model, reconstructed meshes from adjacent video frames often exhibit variations in topology, such as differing numbers of vertices and faces, as well as changes in vertex and edge connectivity. To create a valid 4D animation asset with consistent geometry, we propose an iterative pairwise registration approach. For two consecutive meshes after refinement, \(\widetilde{\mathcal{M}}_{\text{refine}, t}\) and \(\widetilde{\mathcal{M}}_{\text{refine}, t+1}\), we begin by treating \(\widetilde{\mathcal{M}}_{\text{refine}, t}\) as a rigid body. We optimize its global transformation (rotation, translation, and scale) to align it with \(\widetilde{\mathcal{M}}_{\text{refine}, t+1}\) using Chamfer distance and a differentiable rendering loss. After achieving global registration, we refine the alignment using the preconditioned optimization proposed by ~\cite{nicolet2021large}, which allows for fast and smooth convergence in the inverse reconstruction process. However, the differentiable rendering loss from ~\cite{nicolet2021large} can struggle with regions undergoing large deformations or containing long, thin structures. To improve registration in these areas, we incorporate the As-Rigid-As-Possible (ARAP) constraint ~\cite{igarashi2005rigid} alongside the Chamfer distance for enhanced local registration. Through these global and local alignment steps, we obtain a deformed version \(\widetilde{\mathcal{M}}'_{t}\) (with vertices \(\mathcal{V}'_t\) and faces \(\mathcal{F}'_t\)) of \(\widetilde{\mathcal{M}}_{\text{refine}, t}\) (\(\mathcal{V}_t\) and faces \(\mathcal{F}_t\)), ensuring that the number and topology are identical (\(\mathcal{V}'_t = \mathcal{V}_t\), \(\mathcal{F}'_t = \mathcal{F}_t\)) and closely resemble the shape of \(\widetilde{\mathcal{M}}_{\text{refine}, t+1}\). We then replace the initial \(\widetilde{\mathcal{M}}_{\text{refine}, t+1}\) with the deformed \(\widetilde{\mathcal{M}}'_{t}\). In practice, we designate \(\widetilde{\mathcal{M}}_{\text{refine}, 1}\) as the \textit{rest pose} and iteratively apply the registration process across consecutive frames. This ensures consistent geometry across all meshes \(\widetilde{\mathcal{M}}'_{\{1, \dots, T\}}\).

\begin{figure}[!t]
  \centering
   \includegraphics[width=0.8\linewidth]{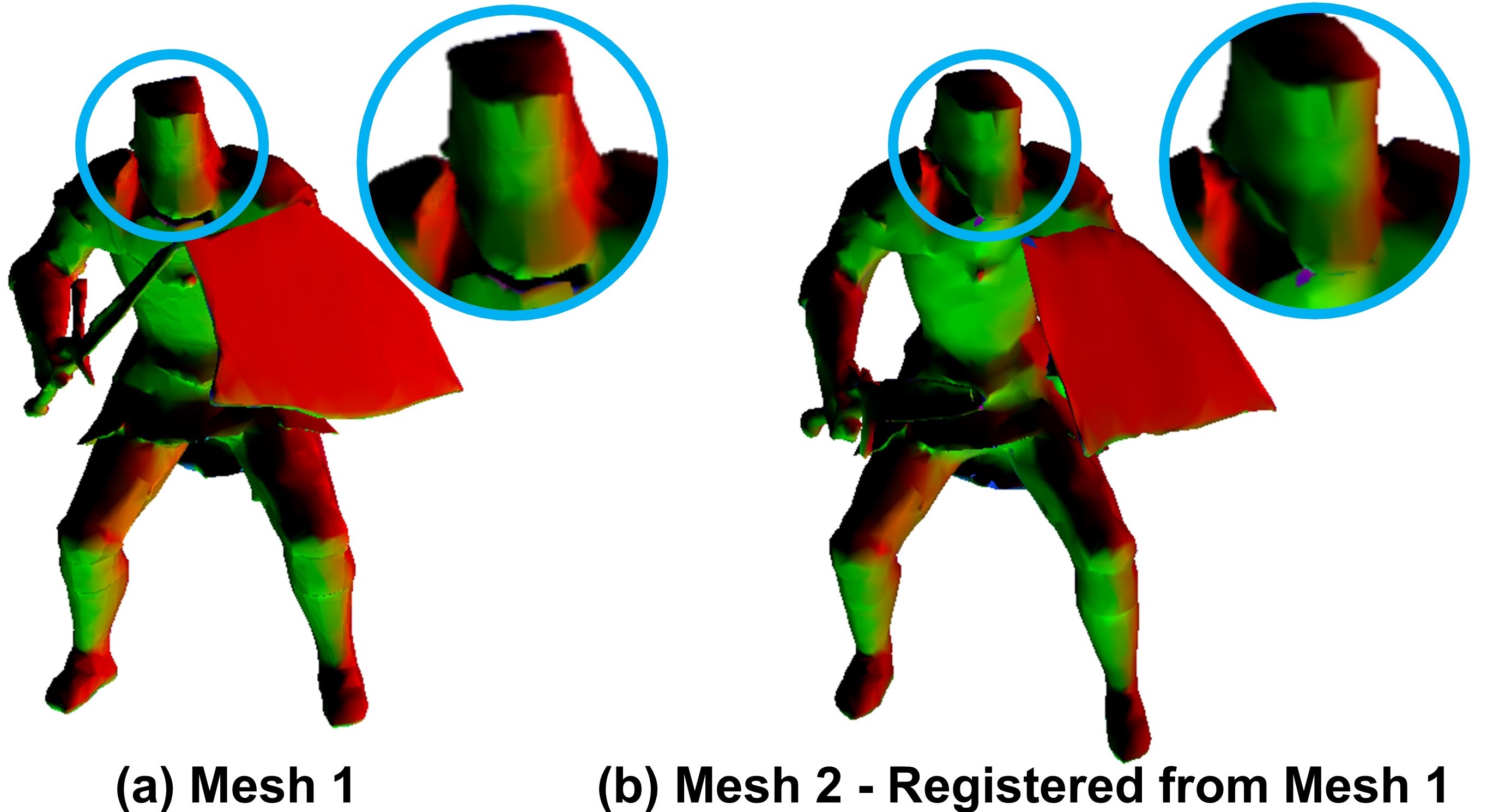}
   \vspace{-5pt}
   \caption{\textbf{Comparison of Vertex Positions Between Original and Registered Meshes.} We compare the vertex correspondences between mesh 1 and mesh 2, where mesh 2 is registered from mesh 1. To visualize these correspondences, we use the vertex normals of mesh 1 to color both meshes, highlighting their alignment.}
   \vspace{-15pt}
   \label{fig:uvalign}
\end{figure}

\subsection{Consistent Texture Map via Weighted Optimization}    
\label{subsec:Consistent_Texture}

After the mesh registration process outlined in Sec.~\ref{subsec:Consistent_Geometry}, we observe that the deformed vertices maintain their correspondence to the initial parts of the object. For instance, vertices representing the left side of the face remain aligned with the left face region (as illustrated in Fig.~\ref{fig:uvalign}). Therefore, we directly inherit the texture map and UV coordinates of \(\widetilde{\mathcal{M}}'_1\) for the subsequent meshes \(\widetilde{\mathcal{M}}'_{\{2, \dots, T\}}\). To refine the texture and minimize artifacts, we further optimize the texture map using multi-view renderings of Gaussian splats generated alongside \(\widetilde{\mathcal{M}}_{\text{refine}, 1}\) (see Sec.~\ref{sec:Preliminaries}), while using a single view under \(\mathcal{C}_{\text{ref}, t}\) for Gaussian splats of the subsequent meshes \(\widetilde{\mathcal{M}}_{\text{refine}, \{2, \dots, T\}}\). This setting is to ensure that all parts of the object are captured, even those that might not be visible in the video. We apply higher weights to views from \(\mathcal{C}_{\text{ref}}\) to ensure accurate alignment with the reference view \(V_{\text{ref}}\). This optimization results in a globally shared texture map \(\mathcal{T}\) for all meshes \(\widetilde{\mathcal{M}}'_{\{1, \dots, T\}}\).

\subsection{Mesh Interpolation and 4D Asset Conversion}
\label{subsec:Mesh Interpolation and 4D Asset Conversion}

Since adjacent video frames, such as $ V_{\text{ref},t} $ and $V_{\text{ref},t+1} $, are typically quite similar, we downsample the time dimension in our practical implementation by selecting every $ i_{\text{th}} $ frame to reconstruct the mesh. This results in a time-reduced mesh sequence. To ensure smooth mesh animation, we linearly interpolate the vertex positions between every two adjacent reconstructed meshes, aligning them with the time resolution of $ V_{\text{ref}} $. Next, we convert the upsampled meshes $ \widetilde{\mathcal{M}}'_{\{1, \dots, T\}} $ into the mesh $ \widetilde{\mathcal{M}}'_{1} $ plus a deformation tensor $ \mathcal{D} \in \mathbb{R}^{T \times N \times 3} $, which stores the vertices' deformation over time. By keyframing these offsets and using the globally shared texture map $ \mathcal{T} $, we can generate a GLTF animation file that is compatible with graphics and game engines.

% =============================================== Experiments ==================================================

\begin{table}[t]
\aboverulesep=0ex
\belowrulesep=0ex
\resizebox{\columnwidth}{!}{\begin{tabular}{@{}c|l|cccc@{}}
\toprule
 & \textbf{Method} & \textbf{CLIP$\uparrow$} & \textbf{LPIPS$\downarrow$} & \textbf{FVD$\downarrow$} & \textbf{DreamSim$\downarrow$} \\ \cmidrule(l){2-6} 
 & Naïve TRELLIS   & 0.8905                  & 0.1597                     & 1342.66                  & 0.1282                        \\
 & DreamMesh4D     & 0.8692                  & 0.1019                     & 914.28                   & 0.0937                        \\ \cmidrule(l){2-6} 
\multirow{-4}{*}{\textbf{Simple}} &
  V2M4 (Ours) &
  \cellcolor[HTML]{EFEFEF}\textbf{0.9259} &
  \cellcolor[HTML]{EFEFEF}\textbf{0.1017} &
  \cellcolor[HTML]{EFEFEF}\textbf{825.59} &
  \cellcolor[HTML]{EFEFEF}\textbf{0.0688} \\ \midrule
 & \textbf{Method} & \textbf{CLIP$\uparrow$} & \textbf{LPIPS$\downarrow$} & \textbf{FVD$\downarrow$} & \textbf{DreamSim$\downarrow$} \\ \cmidrule(l){2-6} 
 & Naïve TRELLIS   & 0.8887                  & 0.1265                     & 1216.19                  & 0.1492                        \\
 & DreamMesh4D     & 0.8256                  & 0.0804                     & 1079.02                  & 0.1850                        \\ \cmidrule(l){2-6} 
\multirow{-4}{*}{\textbf{Complex}} &
  V2M4 (Ours) &
  \cellcolor[HTML]{EFEFEF}\textbf{0.9008} &
  \cellcolor[HTML]{EFEFEF}\textbf{0.0747} &
  \cellcolor[HTML]{EFEFEF}\textbf{666.04} &
  \cellcolor[HTML]{EFEFEF}\textbf{0.1220} \\ \midrule
 & \textbf{Method} & \textbf{CLIP$\uparrow$} & \textbf{LPIPS$\downarrow$} & \textbf{FVD$\downarrow$} & \textbf{DreamSim$\downarrow$} \\ \cmidrule(l){2-6} 
 & Naïve TRELLIS   & 0.8891                  & 0.1352                     & 1014.45                  & 0.1438                        \\
 & DreamMesh4D     & 0.8369                  & 0.0860                     & 855.29                   & 0.1613                        \\ \cmidrule(l){2-6} 
\multirow{-4}{*}{\textbf{All}} &
  V2M4 (Ours) &
  \cellcolor[HTML]{EFEFEF}\textbf{0.9073} &
  \cellcolor[HTML]{EFEFEF}\textbf{0.0817} &
  \cellcolor[HTML]{EFEFEF}\textbf{576.73} &
  \cellcolor[HTML]{EFEFEF}\textbf{0.1082} \\ \bottomrule
\end{tabular}}
\vspace{-5pt}
\caption{\textbf{Quantitative Evaluation of Different Methods.} All metrics are computed using the rendering view of the reconstructed meshes. ``Simple" refers to the Consistent4D data, which consists of simple objects with subtle movements, while ``Complex" refers to data collected online featuring more complex objects and large-scale movements. ``All" represents the combined evaluation on both subsets. The best results are in bold.}
\vspace{-5pt}
\label{Tab:comparisons}
\end{table}

\begin{table}[t]

\centering
\resizebox{0.8\columnwidth}{!}{\begin{tabular}{@{}c|ccc@{}}
\toprule
\textbf{Method} & \textbf{Naïve TRELLIS} & \textbf{DreamMesh4D} & \textbf{V2M4 (Ours)} \\ \midrule
Time      & 40s                    & 3 min                & 60s                  \\ \bottomrule
\end{tabular}}
\vspace{-5pt}
\caption{\textbf{Average Reconstruction Time per Frame.} The displayed time represents the average time taken to reconstruct each frame across an animation sequence. Detailed timing for each step of our method is provided in Appendix~\ref{appendix: implementation details}.}
\vspace{-15pt}
\label{Tab:time}
\end{table}

\begin{figure*}[!t]
  \centering
   \includegraphics[width=\linewidth]{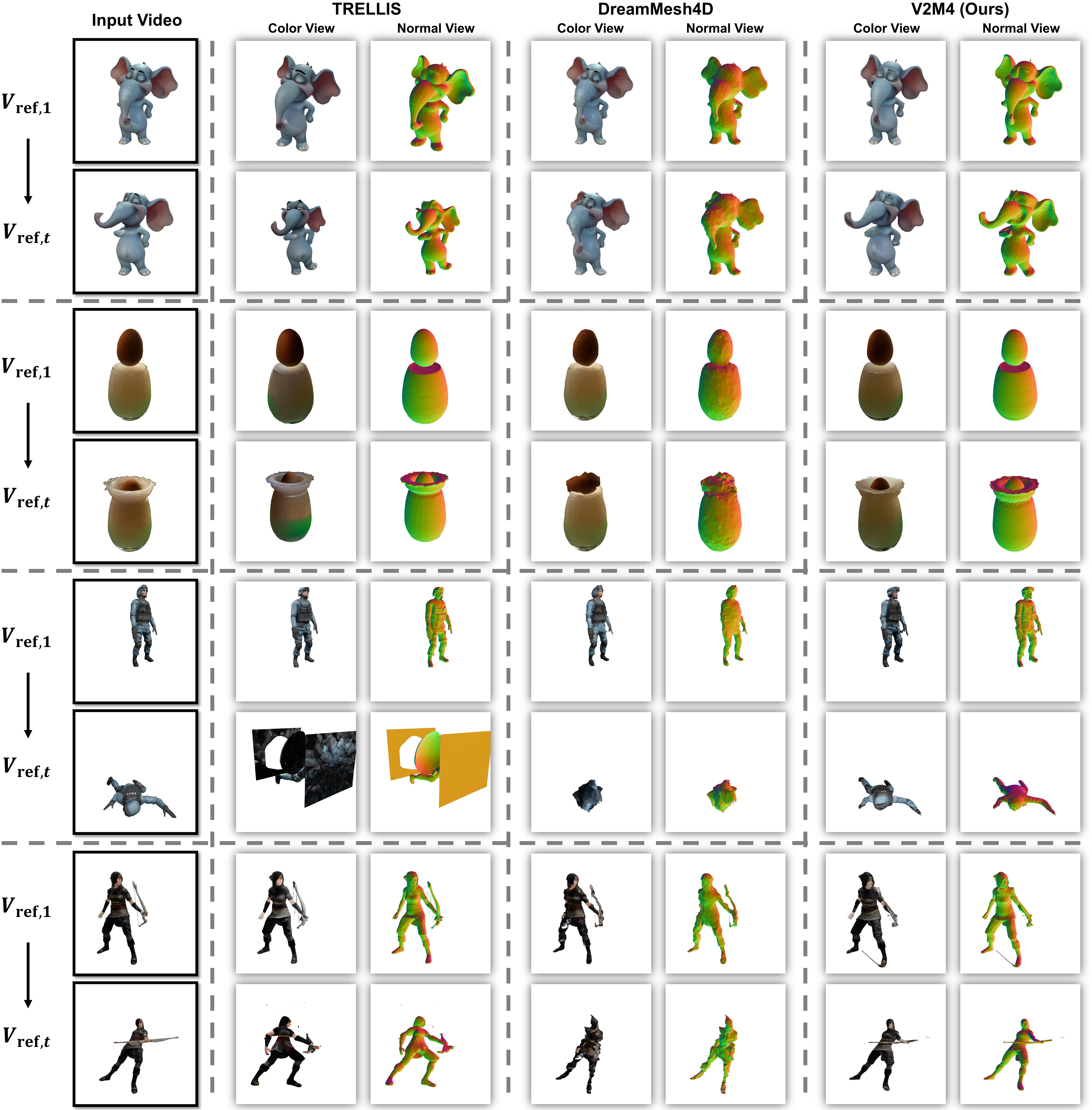}
   \vspace{-20pt}
   \caption{\textbf{Qualitative Evaluation of Different Methods.} We show both the color and normal views of the reconstructed meshes. The first two samples are from the Consistent4D dataset, while the rest are from our collected data. Each sample includes two timestamps for quick comparison. Please zoom in for a clearer view. For more results and animations, see the Supplementary Files.}
   \label{fig:Comparison}
\end{figure*}

\begin{figure}[!t]
  \centering
   \includegraphics[width=\linewidth]{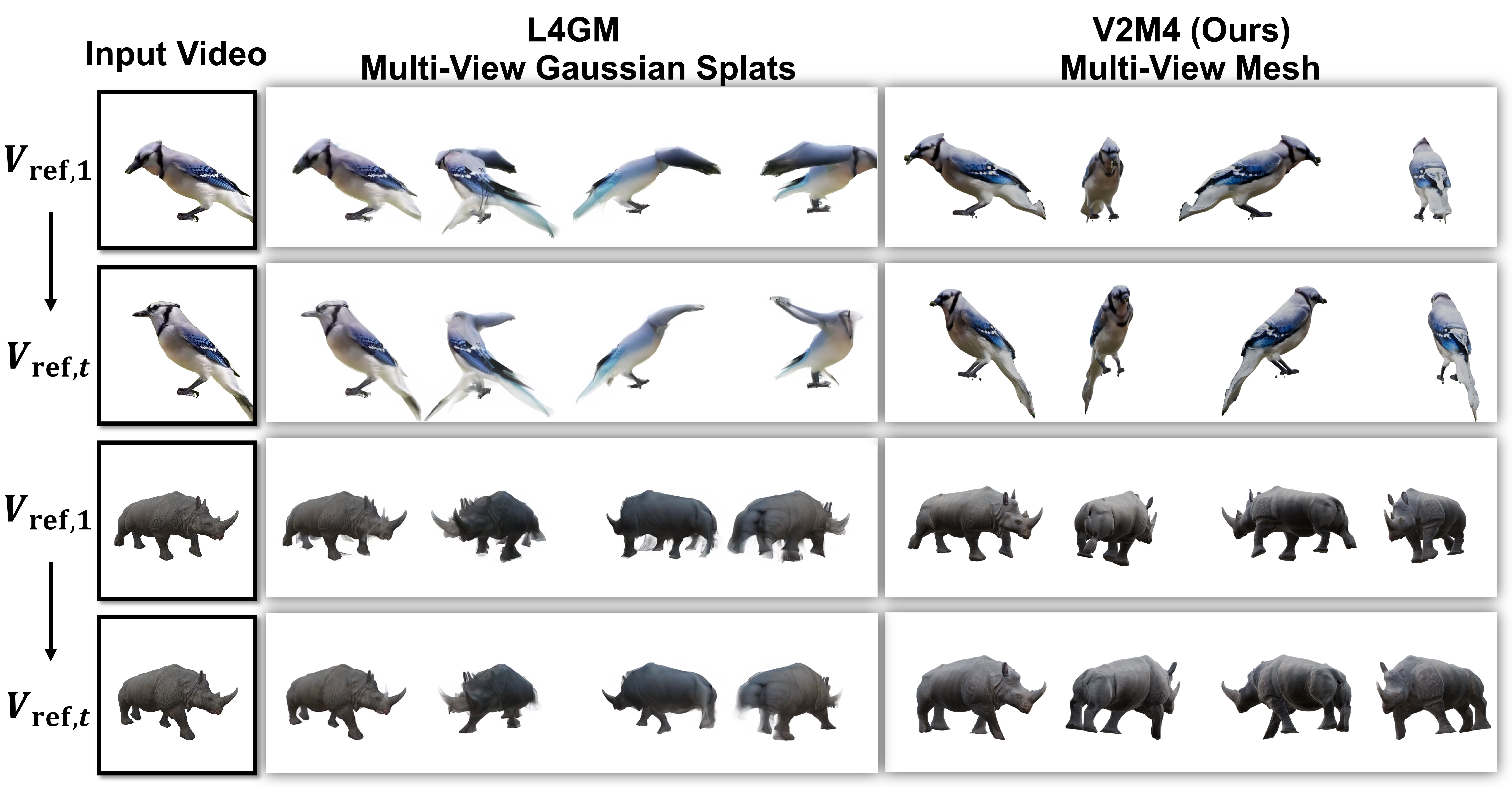}
   \vspace{-20pt}
    \caption{\textbf{Multi-View Object Geometry Comparison with L4GM.} For L4GM, we present the multi-view Gaussian Splats rendering views, while for our method, we display the multi-view rendering views of our mesh. Please zoom in for a clearer view.}
   \label{fig:Comparison2}
   \vspace{-10pt}
\end{figure}

\section{Experiments}
\label{sec:Experiments}

\subsection{Experimental Setup}

Please refer to Appendix~\ref{appendix: implementation details} for implementation details, and to Appendix~\ref{appendix:Ablation Studies}~\&~\ref{append: more visual} for comprehensive ablation studies (including experiments with more advanced 3D generation models such as Hunyuan3D-2.0~\cite{zhao2025hunyuan3d}, TripoSG~\cite{li2025triposg}, and advanced dense stereo reconstruction models like VGGT~\cite{wang2025vggt}), as well as additional visual results.

\vspace{-10pt}
\paragraph{Datasets.} We constructed a benchmark comprising 40 animation videos. Specifically, we first collected 20 animation videos from Consistent4D~\cite{jiang2023consistent4d}, which feature simple object topologies with subtle movements. Additionally, we gathered 20 animation videos from Mixamo~\cite{Mixamo} and Sketchfab~\cite{sketchfab} to include larger-scale object movements and more complex object topologies (with sequences containing more than 200 frames), thus providing a diverse test set for assessing reconstruction performance. Our benchmark is \textbf{much larger} and \textbf{more varied} than those used in existing works~\cite{li2025dreammesh4d, jiang2023consistent4d, zeng2024stag4d}, which evaluate only 7 test samples (each with just 32 frames). Sample frames from our benchmark can be found in Appendix \ref{append: more visual}. Since existing video segmentation methods~\cite{ravi2024sam2} can easily extract foreground object regions, the dataset has been pre-processed to remove backgrounds. This ensures that the evaluation remains focused on reconstruction quality rather than segmentation challenges.

\vspace{-10pt}
\paragraph{Evaluation Metrics.} To evaluate the reconstruction quality, we adopt metrics commonly used in 4D works~\cite{zeng2024stag4d, ren2025l4gm}, including CLIP~\cite{radford2021learning}, LPIPS~\cite{zhang2018perceptual}, and FVD~\cite{unterthiner2018towards}, to assess the visual quality of the reconstructed 4D animation through video-to-rendering evaluation. We also include DreamSim~\cite{fu2023dreamsim} due to its excellent performance in evaluating semantic similarity. All these metrics are calculated between the input video frames and the rendered video frames of the reconstructed mesh with an aligned camera viewpoint. Please see Appendix \ref{appendix: implementation details} for details on the metric settings.

\subsection{Comparisons}
\label{subsec:Comparisons}

% \vspace{-10pt}
\paragraph{Quantitative Evaluation.} Since \awesomemesh directly outputs 4D meshes and most existing 4D works focus on single NeRF or Gaussian Splats outputs, we ensure fair comparisons by evaluating our method against the recent DreamMesh4D~\cite{li2025dreammesh4d}, which outputs both Gaussian Splats and 4D meshes. We also compare our results with the naive TRELLIS~\cite{xiang2024structured}, which reconstructs each video frame individually. Table~\ref{Tab:comparisons} presents the quantitative performance results on our benchmark. For comprehensiveness, we separately display results on Consistent4D data, which features simple topology and subtle movement, and our additional data from Mixamo and Sketchfab, which includes complex topology, intense movement, and longer movement durations. From the results, \awesomemesh outperforms both other methods, achieving superior performance across all metrics. Additionally, we report processing times in Table~\ref{Tab:time}. Thanks to our interpolation design detailed in Section~\ref{subsec:Mesh Interpolation and 4D Asset Conversion}, our method processes each frame in approximately one minute on average, adding little to the naive TRELLIS processing time, while being significantly faster than DreamMesh4D.

\vspace{-10pt}
\paragraph{Qualitative Evaluation.}  Fig.~\ref{fig:Comparison} show qualitative visual results of the reconstructed meshes across different methods on the Consistent4D data and the newly collected data. We present both the rendering color view and normal view of the meshes. From these visuals, \awesomemesh produces mesh results with significantly better appearance and topology than DreamMesh4D, and they are much more aligned with the reference videos. These findings verify that \awesomemesh is effective and generalizable, capable of reconstructing high-quality, usable 4D mesh animations from monocular videos.

Additionally, to further highlight the advantages of basing the 4D reconstruction task on native 3D mesh models, we qualitatively compare our method with a current powerful 4D Gaussian Splatting reconstruction method, L4GM~\cite{ren2025l4gm}, which explicitly trains a multi-view Gaussian Splatting network. In Fig.~\ref{fig:Comparison2}, we display the multi-view rendering views of our mesh and Gaussian Splats for several cases. We observe that \awesomemesh preserves more accurate model geometry than L4GM, underscoring the promise of using native 3D mesh generation models for 4D tasks.

% =============================================== Limitations and Outlook ==================================================

\subsection{Limitations}  
\label{sec:Limitations and Outlook}  

While our method effectively reconstructs 4D mesh animations from input videos in many scenarios, it has several limitations. First, since our approach is based on TRELLIS, our method may suffer from performance issues when TRELLIS outputs poor 3D reconstructions (see further analysis in Appendix~\ref{appendix:Ablation Studies}). Second, artifacts can occur when reconstructing animations with topology changes, as we enforce topology consistency by registering the first mesh to subsequent frames. Third, using a fixed sampling rate, as discussed in Section~\ref{subsec:Mesh Interpolation and 4D Asset Conversion}, can be inefficient due to variable motion dynamics—sometimes moving slowly and other times rapidly. Future work could benefit from incorporating adaptive sampling techniques.

% =============================================== Conclusion ==================================================
\section{Conclusion}
\label{sec:Conclusion}

In this work, we propose a 4D reconstruction method capable of directly generating usable 4D mesh animation assets. Our method is built upon a 3D mesh generation model and addresses key challenges, including misalignment between the reconstructed mesh and the input video in terms of pose and appearance, as well as inconsistencies in geometry and texture throughout the animation. Our approach successfully produces high-quality mesh animation files that are compatible with graphics and game software. We believe this work offers a promising solution by enabling direct 4D mesh generation.

\clearpage

\section*{Acknowledgements}
This work was supported by funding from King Abdullah University of Science and Technology (KAUST) — Center of Excellence for Generative AI, under award number 5940.

{
    \small
    \bibliographystyle{ieeenat_fullname}
    \bibliography{main}
}

\clearpage

\appendix

\section{Implementation Details}
\label{appendix: implementation details}

In Sec.~\ref{sec:Method} of the main paper, we described the design of each component in our framework. These components are sequentially integrated according to the overall pipeline shown in Fig.~\ref{fig:framework} of the main paper. Below, we provide additional implementation details for each component of our framework.

\paragraph{Camera Search.}
In the camera search workflow, we set the number of initially sampled large-scale camera poses \( P_0 \) to 2000. The \(\mathrm{yaw}\) and \(\mathrm{pitch}\) angles are sampled using an equal-area distribution around the sphere. The \(\mathrm{radius}\) is sampled with steps following a square root distribution within the range [1.0, 5.0], and \(\mathrm{lookat}_{x,y,z}\) are uniformly sampled within the range [-1.0, 1.0] for all three axes. We select the top \( n = 7 \) camera poses, along with the reference video frame, to input to DUSt3R. The scoring function between the rendering view and the reference view is a combination of the DreamSim loss~\cite{fu2023dreamsim} and the foreground mask area loss, with respective weights of 1 and 0.1.

In the utilization of the dense stereo reconstruction model, we obtain the ground truth point clouds of \(\widetilde{\mathcal{M}}_t\) during the rasterization process using Nvdiffrast~\cite{Laine2020diffrast}. To speed up optimization and filter out possible outliers in the point clouds, we retain only 5\% points of both the ground truth point clouds \(\mathrm{PC}_{\text{rend}, {\{1, \dots, n\}}}\) and the predicted ones \(\overline{\mathrm{PC}}_{\text{rend}, {\{1, \dots, n\}}}\). We then align these two sets of point clouds using Chamfer loss to optimize the global transformation (rotation, scale, and translation) over 2000 iterations. To prevent the alignment from becoming trapped in local minima, such as an object being flipped along its vertical axis, we manually flip the point clouds along the axis starting at the 1000th iteration. We then compare the Chamfer loss before and after flipping to select the best alignment. After aligning the point cloud sets, we optimize the camera pose for 500 iterations using MSE loss between the 3D point clouds of \(\mathrm{PC}_{\text{ref}}\) and the 2D pixel positions in \(V_{\text{ref}, t}\).

We select the top \( K = 199 \) camera poses from \( P_0 \), along with the camera pose from DUSt3R, for the PSO algorithm. We set the number of iterations for PSO to 25. For the subsequent gradient descent optimization, we set the iterations to 300 and use only the MSE loss between the foreground mask area of the rendering view and the reference video frame.

\paragraph{Mesh Appearance Refinement.} We enable optimization of the negative condition embedding in TRELLIS. At each inference step of the generative model (conditioned on this embedding), we decode the output into a mesh. By computing the visual alignment between the rendering view of the reconstructed mesh and Gaussian Splats, we backpropagate the gradient through differentiable rendering to the negative condition embedding in TRELLIS. We use a combination of DreamSim, LPIPS, and MSE losses (each with a weight of 1), along with an MSE regularization term weighted at 0.2 to compare with the embedding's initial value, preventing distortion of the reconstructed result. This optimization is applied only during the latter part of the flow model (\(0.6 \geq t \geq 0\), where \(t\) represents the timestep of the flow model). We begin the optimization at iteration 5 and gradually increase by 1 every 5 timesteps of the flow model.

\paragraph{Topology Consistency.} For global alignment, we use Chamfer loss and L1 loss between the rendering views of the two meshes, with equal weighting for both losses. The optimization is performed over 500 iterations, using 20 randomly selected views around the mesh for the rendering loss.

For local alignment, we set the iterations to 1000 and use 50 views for the rendering view L1 loss. In addition to Chamfer loss, we incorporate ARAP loss, Face Area Consistency loss (which penalizes face area changes during mesh deformation), and Edge Length Consistency loss (which penalizes edge length changes). Due to the small values of Face Area Consistency loss and Edge Length Consistency loss, we assign them weights of \(1e^6\) and \(1e^2\), respectively. All other losses are assigned a weight of 1.

\paragraph{Texture Consistency.} For the first mesh, we select 100 random views around it, and for the subsequent meshes, we use only their rendering views under \(\mathcal{C}_\text{ref}\). The global texture map is optimized based on all these rendering views over 2500 iterations, using both L1 loss and total variation loss to ensure natural smoothness.

\paragraph{Mesh Interpolation.} We set the frame interpolation to 5 for the complex data collected online due to their high FPS, and to 3 for the Consistent4D data, which has a relatively low FPS.

\begin{table}[t]
\renewcommand{\arraystretch}{1.3}
\resizebox{\columnwidth}{!}{\begin{tabular}{c|c|ccc}
\hline
\multirow{2}{*}{\textbf{Total}} & \multirow{2}{*}{Mesh Gen} & \multicolumn{3}{c}{Camera Pose Search}                                                          \\ \cline{3-5} 
                                &                           & \multicolumn{1}{c|}{DUSt3R}            & \multicolumn{1}{c|}{PSO}         & \multicolumn{1}{c}{SGD} \\ \hline
 \cellcolor[HTML]{CECECE}\textbf{57.7s} & 1.5s &  \multicolumn{1}{c|}{13.9s} & \multicolumn{1}{c|}{11.1s} &  0.5s \\ \hline
Mesh Refine                     & Mesh Regist               & \multicolumn{1}{c|}{Texture Map Optim} & \multicolumn{1}{c|}{Mesh Interp} & 4D Asset Convert        \\ \hline
 14.1s &  15.6s & \multicolumn{1}{c|}{0.3s} & \multicolumn{1}{c|}{0.6s} & 0.1s \\ \hline
\end{tabular}
}
\caption{Detailed average runtime per frame for each component of our framework.}
\label{Tab:Runtime}
\end{table}

\paragraph{Metric Calculation Settings.} For accurate visual similarity evaluation, we use the ``ViT-bigG-14" model provided by OpenCLIP~\cite{ilharco_gabriel_2021_5143773}, trained on the LAION-2B~\cite{schuhmann2022laion} dataset, to calculate the CLIP score. For FVD calculation, we use StyleGAN-V~\cite{stylegan_v}. For the LPIPS metric calculation, we use the VGG model. Since the FVD metric requires input videos to have the same number of frames, for reconstruction results on our additionally collected long animation videos, we split the rendering video into subsequences of 32-frame videos and calculate FVD on all of them. For the final subsequence that has fewer than 32 frames, we exclude it from the calculation.

All experiments in this paper were conducted on a single A100 GPU. Table~\ref{Tab:Runtime} provides detailed runtime information for each framework component as a supplement to Table~\ref{Tab:time} in the main paper.

\section{Ablation Studies}
\label{appendix:Ablation Studies}

\subsection{Quantitative Ablation Studies}

\paragraph{Ablation Study on Different Base 3D Generation Models.} In Table~\ref{Tab:hunyuan}, we replace our base 3D generator, TRELLIS, with several contemporary models, including Hunyuan3D-2.0~\cite{zhao2025hunyuan3d}, TripoSG~\cite{li2025triposg}, and CraftsMan3D~\cite{li2025craftsman3d}. The results demonstrate that using more advanced 3D generators (such as Hunyuan3D-2.0 and TripoSG) leads to corresponding improvements in performance, illustrating the extensibility of our method alongside ongoing advancements in 3D generation techniques.

\begin{table}[t]
\renewcommand{\arraystretch}{1.3}
\resizebox{\columnwidth}{!}{\begin{tabular}{c|l|cccc}
\toprule
\textbf{Dataset} & \multicolumn{1}{c|}{\textbf{Method}} & \textbf{CLIP$\uparrow$} & \textbf{LPIPS$\downarrow$} & \textbf{FVD$\downarrow$} & \textbf{DreamSim$\downarrow$} \\
\midrule
\multirow{8}{*}{Simple}  
  & Naïve TRELLIS        & 0.8905 & 0.1597 & 1342.66 & 0.1282 \\
  & Naïve CraftsMan3D & 0.9177 & 0.1702 & 1185.47 & 0.0924 \\
  & Naïve TripoSG  & \cellcolor[HTML]{CECECE}\textbf{0.9259} & 0.1534 & 973.16 & 0.0657 \\
  & Naïve Hunyuan3D 2.0  & 0.9185 & \cellcolor[HTML]{CECECE}\textbf{0.1471} & \cellcolor[HTML]{CECECE}\textbf{851.01} & \cellcolor[HTML]{CECECE}\textbf{0.0647} \\ \cline{2-6} 
  & V2M4 (TRELLIS)       & 0.9259 & 0.1017 & 825.59 & 0.0688 \\
  & V2M4 (CraftsMan3D)       & 0.9212 & 0.1051 & 960.63 & 0.0772 \\
  & V2M4 (TripoSG)       & \cellcolor[HTML]{CECECE}\textbf{0.9286} & 0.0971 & 758.25 & 0.0652 \\
  & V2M4 (Hunyuan3D 2.0)  & \cellcolor[HTML]{CECECE}\textbf{0.9286} & \cellcolor[HTML]{CECECE}\textbf{0.0885} & \cellcolor[HTML]{CECECE}\textbf{691.48} & \cellcolor[HTML]{CECECE}\textbf{0.0634} \\
\midrule
\multirow{8}{*}{Complex} 
  & Naïve TRELLIS        & 0.8887 & 0.1265 & 1216.19 & 0.1492 \\
  & Naïve CraftsMan3D & 0.9060 & 0.1350 & 1162.00 & 0.1027 \\
  & Naïve TripoSG  & \cellcolor[HTML]{CECECE}\textbf{0.9201} & 0.1404 & 1023.88 & \cellcolor[HTML]{CECECE}\textbf{0.0891} \\
  & Naïve Hunyuan3D 2.0  & 0.9097 & \cellcolor[HTML]{CECECE}\textbf{0.1185} & \cellcolor[HTML]{CECECE}\textbf{1022.81} & 0.0962 \\ \cline{2-6} 
  & V2M4 (TRELLIS)       & 0.9008 & 0.0747 & 666.04 & 0.1220 \\
  & V2M4 (CraftsMan3D)       & 0.9268  & 0.0709 & 558.34 & 0.1046 \\
  & V2M4 (TripoSG)       & \cellcolor[HTML]{CECECE}\textbf{0.9359}  & 0.0608 & 433.38 & \cellcolor[HTML]{CECECE}\textbf{0.0873} \\
  & V2M4 (Hunyuan3D 2.0)   & 0.9192 & \cellcolor[HTML]{CECECE}\textbf{0.0605} & \cellcolor[HTML]{CECECE}\textbf{415.53} & 0.0990 \\
\midrule
\multirow{8}{*}{All}     
  & Naïve TRELLIS        & 0.8891 & 0.1352 & 1014.45 & 0.1438 \\
  & Naïve CraftsMan3D & 0.9098 & 0.1446 & 970.24 & 0.0999 \\
  & Naïve TripoSG  & \cellcolor[HTML]{CECECE}\textbf{0.9222} & 0.1436 & 859.04 & \cellcolor[HTML]{CECECE}\textbf{0.0833} \\
  & Naïve Hunyuan3D 2.0  & 0.9144 & \cellcolor[HTML]{CECECE}\textbf{0.1284} & \cellcolor[HTML]{CECECE}\textbf{796.98} & 0.0853 \\ \cline{2-6} 
  & V2M4 (TRELLIS)       & 0.9073 & 0.0817 & 576.73 & 0.1082 \\
  & V2M4 (CraftsMan3D)       & 0.9270 & 0.0795 & 524.00 & 0.0977 \\
  & V2M4 (TripoSG)       & \cellcolor[HTML]{CECECE}\textbf{0.9359} & 0.0698 & 401.71 & \cellcolor[HTML]{CECECE}\textbf{0.0818} \\
  & V2M4 (Hunyuan3D 2.0)  & 0.9224 & \cellcolor[HTML]{CECECE}\textbf{0.0674} & \cellcolor[HTML]{CECECE}\textbf{377.14} & 0.0901 \\
\bottomrule
\end{tabular}
}
\caption{Impact of different base 3D generators on the overall performance.}
\label{Tab:hunyuan}
\end{table}

\begin{table}[t]
\renewcommand{\arraystretch}{1.3}
\resizebox{\columnwidth}{!}{\begin{tabular}{@{}c|c|cccc@{}}
\toprule
\textbf{Base Model} & \textbf{Strategy} & \textbf{CLIP$\uparrow$} & \textbf{LPIPS$\downarrow$} & \textbf{FVD$\downarrow$} & \textbf{DreamSim$\downarrow$} \\ \midrule
\multirow{6}{*}{TRELLIS} & Final                    & \cellcolor[HTML]{CECECE}\textbf{0.9259} & 0.1017 & 825.59 & 0.0688 \\
                         & w/o PSO                  & 0.8855 & 0.1158 & 1090.86 & 0.1184 \\
                         & w/o DUSt3R               & 0.9236 & 0.1046 & \cellcolor[HTML]{CECECE}\textbf{756.29} & 0.0713 \\
                         & w/o SGD                  & 0.9146 & 0.1269 & 1086.14 & 0.0861 \\
                         & w/o Mesh Refinement      & 0.9028 & 0.1103 & 1023.60 & 0.1032 \\
                         & Replace DUSt3R with VGGT  & 0.9230 & \cellcolor[HTML]{CECECE}\textbf{0.1006} & 821.22 & \cellcolor[HTML]{CECECE}\textbf{0.0631} \\ \midrule
CraftsMan3D              & Replace DUSt3R with VGGT  & 0.9207 & 0.1046 & 887.53 & 0.0753 \\
TripoSG                  & Replace DUSt3R with VGGT  & 0.9250 & 0.0960 & 730.13 & 0.0757 \\
Hunyuan3D-2.0            & Replace DUSt3R with VGGT  & \cellcolor[HTML]{CECECE}\textbf{0.9279} & \cellcolor[HTML]{CECECE}\textbf{0.0879} & \cellcolor[HTML]{CECECE}\textbf{727.24} & \cellcolor[HTML]{CECECE}\textbf{0.0636} \\ \bottomrule
\end{tabular}
}
\caption{Ablation study about the impact of key components in our framework.}
\label{Tab:ablation}
\end{table}

\begin{table}[t]
\renewcommand{\arraystretch}{1.2}
\resizebox{\columnwidth}{!}{\begin{tabular}{@{}c|c|cccc@{}}
\toprule
\textbf{Parameter} & \textbf{Value} & \textbf{CLIP$\uparrow$} & \textbf{LPIPS$\downarrow$} & \textbf{FVD$\downarrow$} & \textbf{DreamSim$\downarrow$} \\ \midrule
\multirow{3}{*}{\begin{tabular}[c]{@{}c@{}}DreamSim\\ (Camera Search)\end{tabular}} 
                    & 0.1   & 0.9176 & 0.1058 & \cellcolor[HTML]{CECECE}\textbf{824.20} & 0.0784 \\
                    & 1   & \cellcolor[HTML]{CECECE}\textbf{0.9259} & \cellcolor[HTML]{CECECE}\textbf{0.1017} & 825.59 & \cellcolor[HTML]{CECECE}\textbf{0.0688} \\
                    & 10  & 0.9234 & 0.1037 & 853.18 & 0.0726 \\ \midrule
\multirow{3}{*}{\begin{tabular}[c]{@{}c@{}}DreamSim\\ (Mesh Refinement)\end{tabular}}   
                    & 0.1   & 0.9129 & 0.1054 & 876.41 & 0.0911 \\
                    & 1  & \cellcolor[HTML]{CECECE}\textbf{0.9259} & \cellcolor[HTML]{CECECE}\textbf{0.1017} & \cellcolor[HTML]{CECECE}\textbf{825.59} & \cellcolor[HTML]{CECECE}\textbf{0.0688} \\
                    & 10  & 0.9226 & 0.1078 & 932.72 & 0.0750 \\ \midrule
\multirow{3}{*}{ARAP}                                                                   
                    & 0.1   & 0.9200 & 0.1046 & \cellcolor[HTML]{CECECE}\textbf{814.31} & 0.0727 \\
                    & 1  & \cellcolor[HTML]{CECECE}\textbf{0.9259} & \cellcolor[HTML]{CECECE}\textbf{0.1017} & 825.59 & \cellcolor[HTML]{CECECE}\textbf{0.0688} \\
                    & 10  & 0.9258 & 0.1058 & 928.54 & 0.0744 \\ \bottomrule
\end{tabular}
}
\caption{Impact of varying DreamSim and ARAP loss weights.}
\label{Tab:param impact}
\end{table}

\begin{table}[t]
    \renewcommand{\arraystretch}{1.}
    \centering
    \resizebox{0.7\linewidth}{!}{\begin{tabular}{@{}c|cccc@{}}
    \toprule
    \textbf{Noise} & \textbf{CLIP$\uparrow$} & \textbf{LPIPS$\downarrow$} & \textbf{FVD$\downarrow$} & \textbf{DreamSim$\downarrow$} \\ \midrule
    Original  & 0.9259 & 0.1017 & 825.59 & 0.0688 \\ \midrule
    0.1$\times$   & 0.9289 & 0.1022 & 942.99 & 0.0686 \\
    1$\times$  & 0.9259 & 0.1077 & 947.57 & 0.0771 \\
    10$\times$  & 0.8016 & 0.1694 & 2410.03 & 0.2577 \\ \bottomrule
    \end{tabular}}
    \caption{Noise impact.}
    \label{Tab:noise}
\end{table}

\begin{figure*}[!t]
  \centering
   \includegraphics[width=0.8\linewidth]{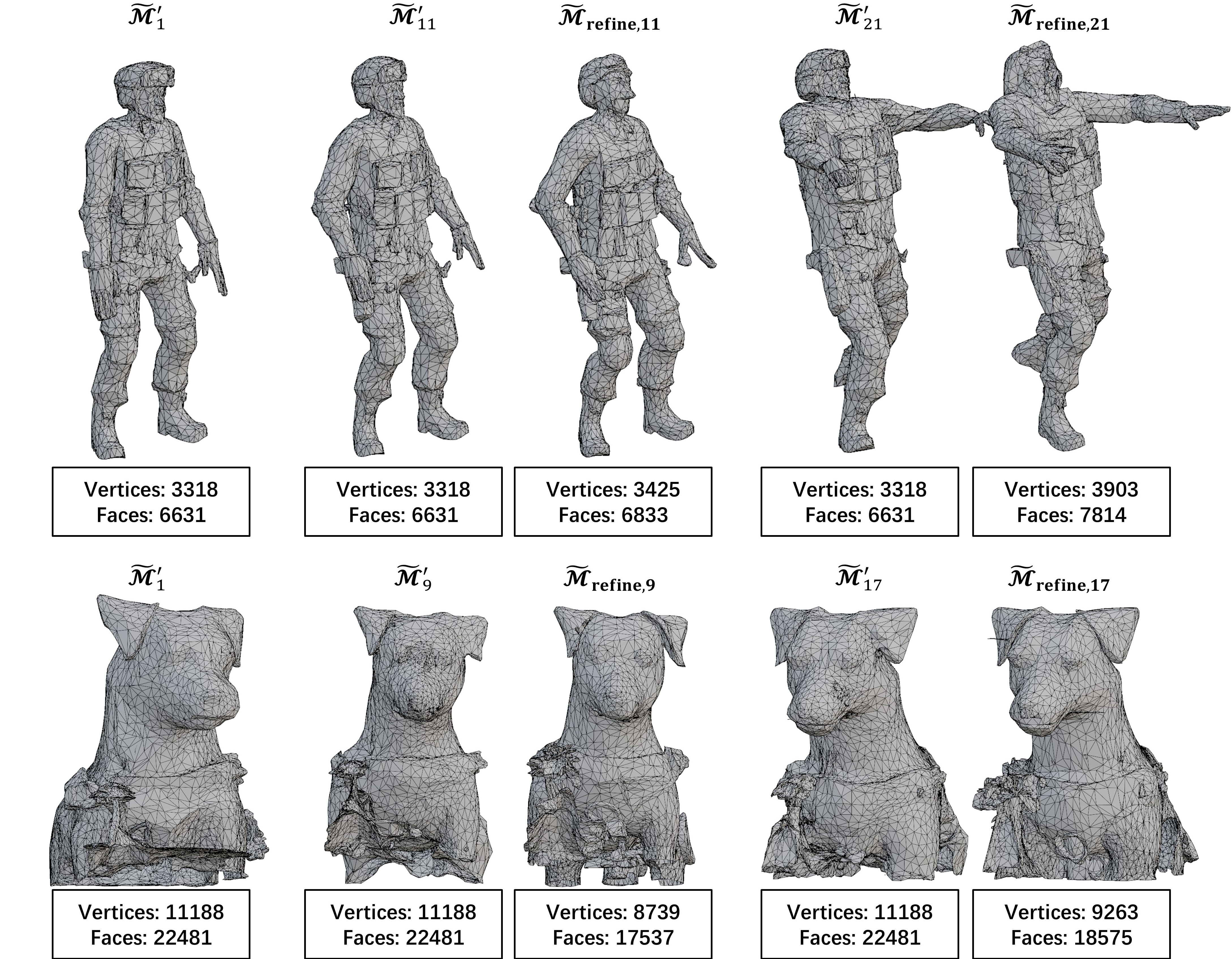}
   \caption{\textbf{Topological Consistency Between Meshes.} \(\widetilde{\mathcal{M}}'_{t}\) represents the registered meshes derived from \(\widetilde{\mathcal{M}}'_{1}\), while \(\widetilde{\mathcal{M}}_{\text{refine},t}\) denotes the original reconstructed meshes. The number of vertices and faces is displayed for clearer comparison.}
   \label{fig:Ablate-Topology}
\end{figure*}

\begin{figure}[!t]
  \centering
   \includegraphics[width=\linewidth]{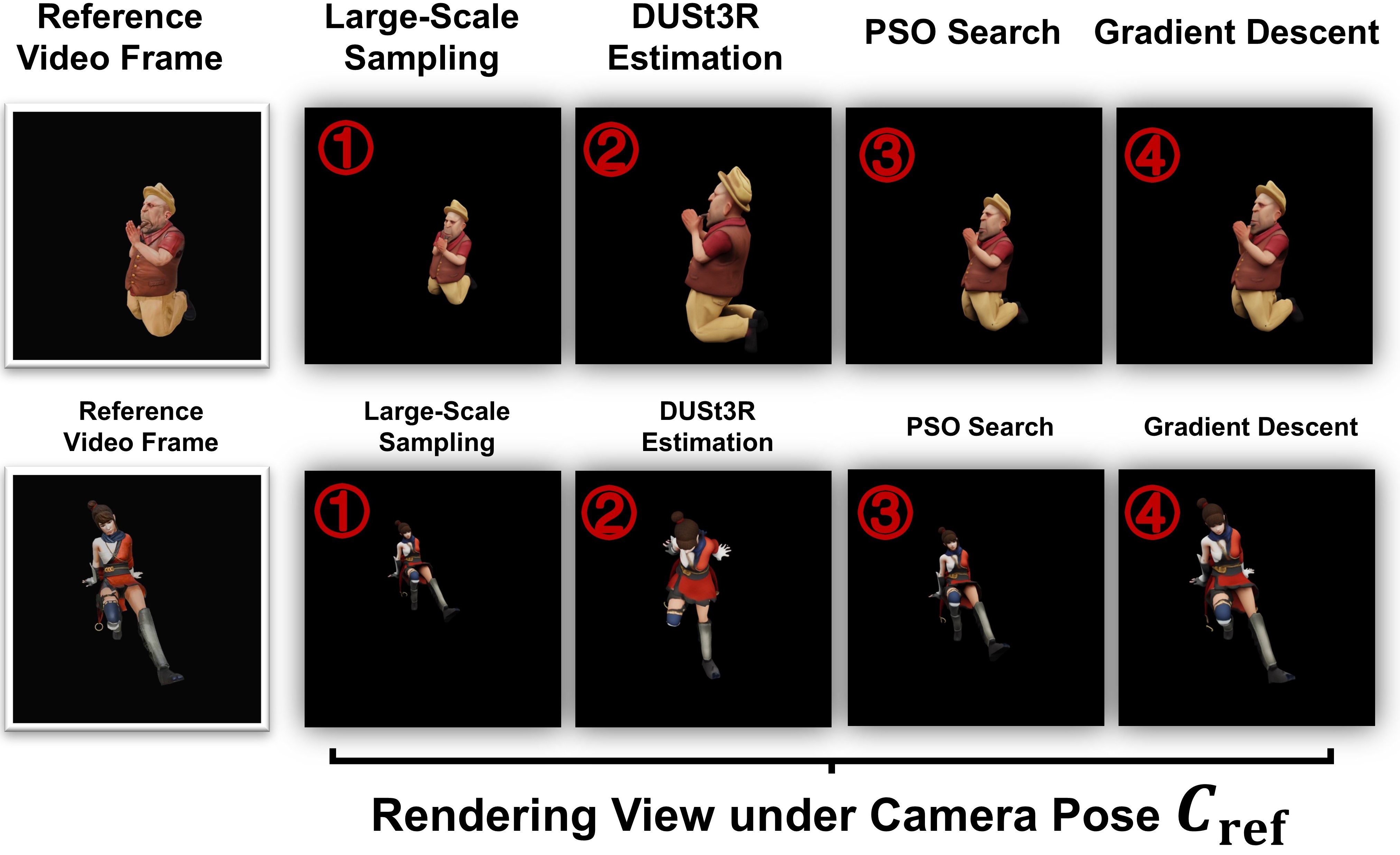}
   \caption{\textbf{Performance of the Camera Search Design.} We present intermediate results from each phase of our camera search workflow. The order of the intermediate results is highlighted with red index numbers.}
   \label{fig:Ablate-Camera}
\end{figure}

\vspace{-10pt}
\paragraph{Ablation of Components and Framework Complexity Analysis.} Table~\ref{Tab:ablation} presents an ablation study of the key components in our framework, confirming the effectiveness of each procedure. We also evaluate the impact of replacing the default dense stereo reconstruction model DUSt3R with the more advanced VGGT~\cite{wang2025vggt}, demonstrating additional performance improvements. 

Regarding theoretical complexity, our framework exhibits linear time complexity $\mathcal{O}(T)$ with respect to the number of frames $T$, while maintaining almost constant memory usage $\mathcal{O}(1)$.

\vspace{-10pt}
\paragraph{Quantitative Parameter Tuning for DreamSim and ARAP.} As outlined in Appendix~\ref{appendix: implementation details}, during the optimization process, we assign only a loss weight factor for DreamSim and ARAP losses, which are set to 1 by default. Table~\ref{Tab:param impact} illustrates the performance impact of varying these parameters.

\vspace{-10pt}
\paragraph{Robustness to Base Mesh Quality.} As discussed in the limitations section of the main paper, the quality of the base mesh significantly impacts the results. Table~\ref{Tab:hunyuan} shows that higher-quality meshes produced by advanced 3D generators (e.g., Hunyuan3D-2.0 and TripoSG) yield improved robustness. Additionally, Table~\ref{Tab:noise} provides a detailed analysis of robustness by evaluating the effects of injecting varying levels of Gaussian noise during mesh generation.

\subsection{Qualitative Ablation Studies}

\paragraph{Topology Consistency.} In Fig.~\ref{fig:Ablate-Topology}, we present both the registered meshes and the original mesh throughout the animation. The original mesh successfully reconstructs to match subsequent timestamp meshes, demonstrating the effectiveness of our design in Sec.~\ref{subsec:Consistent_Geometry}.

\paragraph{Camera Search.} In Fig.~\ref{fig:Ablate-Camera}, we display the rendering views under the identified camera poses at different phases of the camera search workflow described in Sec. \ref{subsec:Semantic Camera Alignment}. Specifically, we show the top-1 camera view after extensive camera pose sampling, the camera view obtained from DUSt3R estimation, the camera view after the Particle Swarm Optimization (PSO) search, and the final camera view following gradient descent refinement (See details in Algorithm~\ref{alg:camera_alignment}). The results demonstrate that our camera search method is both effective and robust, successfully finding the camera pose that aligns with the reference video frame. This alignment subsequently supports accurate mesh reposing.

\begin{figure}[!t]
  \centering
   \includegraphics[width=\linewidth]{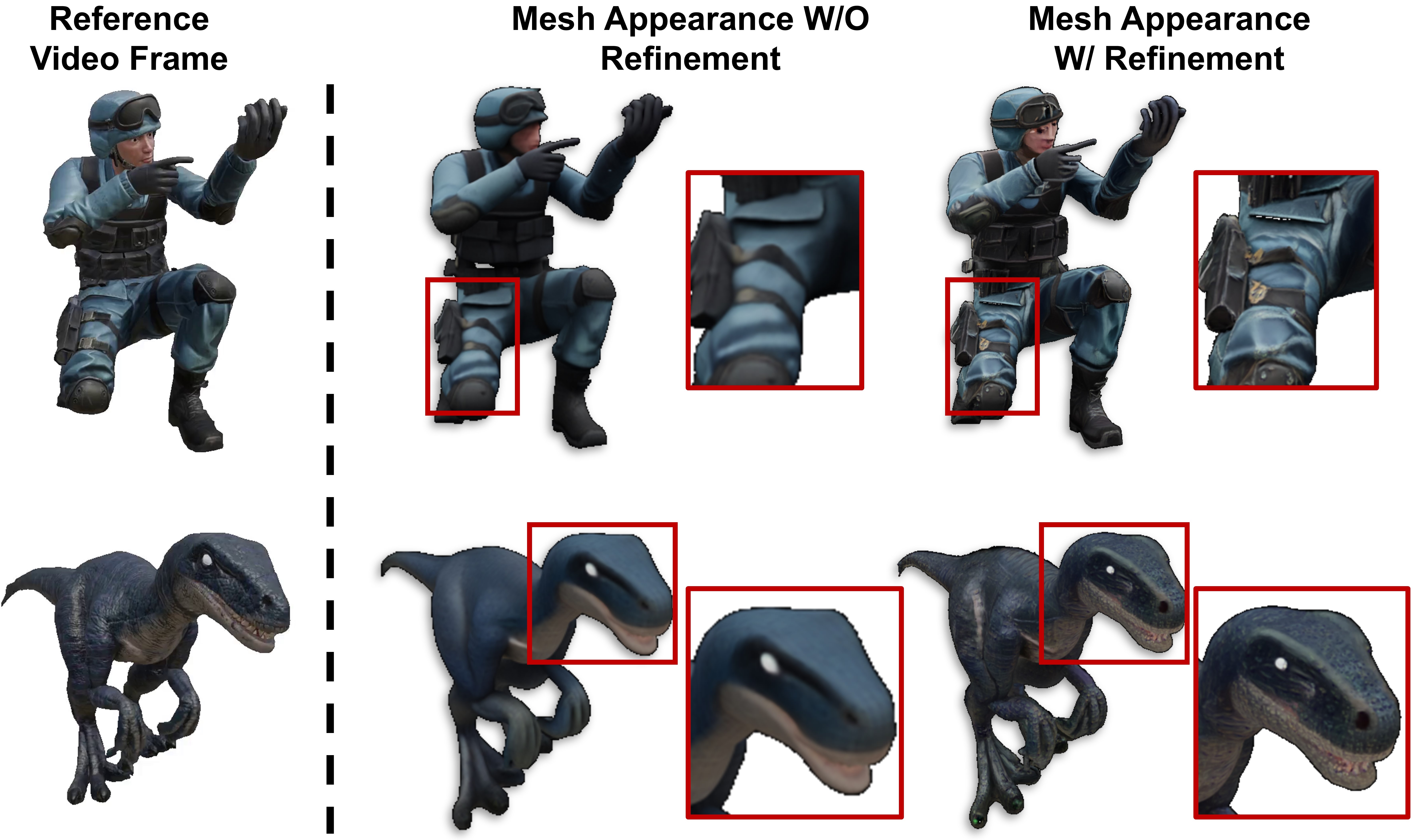}
   \caption{\textbf{Performance of the Mesh Appearance Refinement.} We present the mesh appearance before and after applying our mesh appearance refinement technique. Specific parts are enlarged for better comparison. Please zoom in for a clearer view.}
   \label{fig:Ablate-Appearance}
\end{figure}

\begin{figure}[!t]
  \centering
   \includegraphics[width=\linewidth]{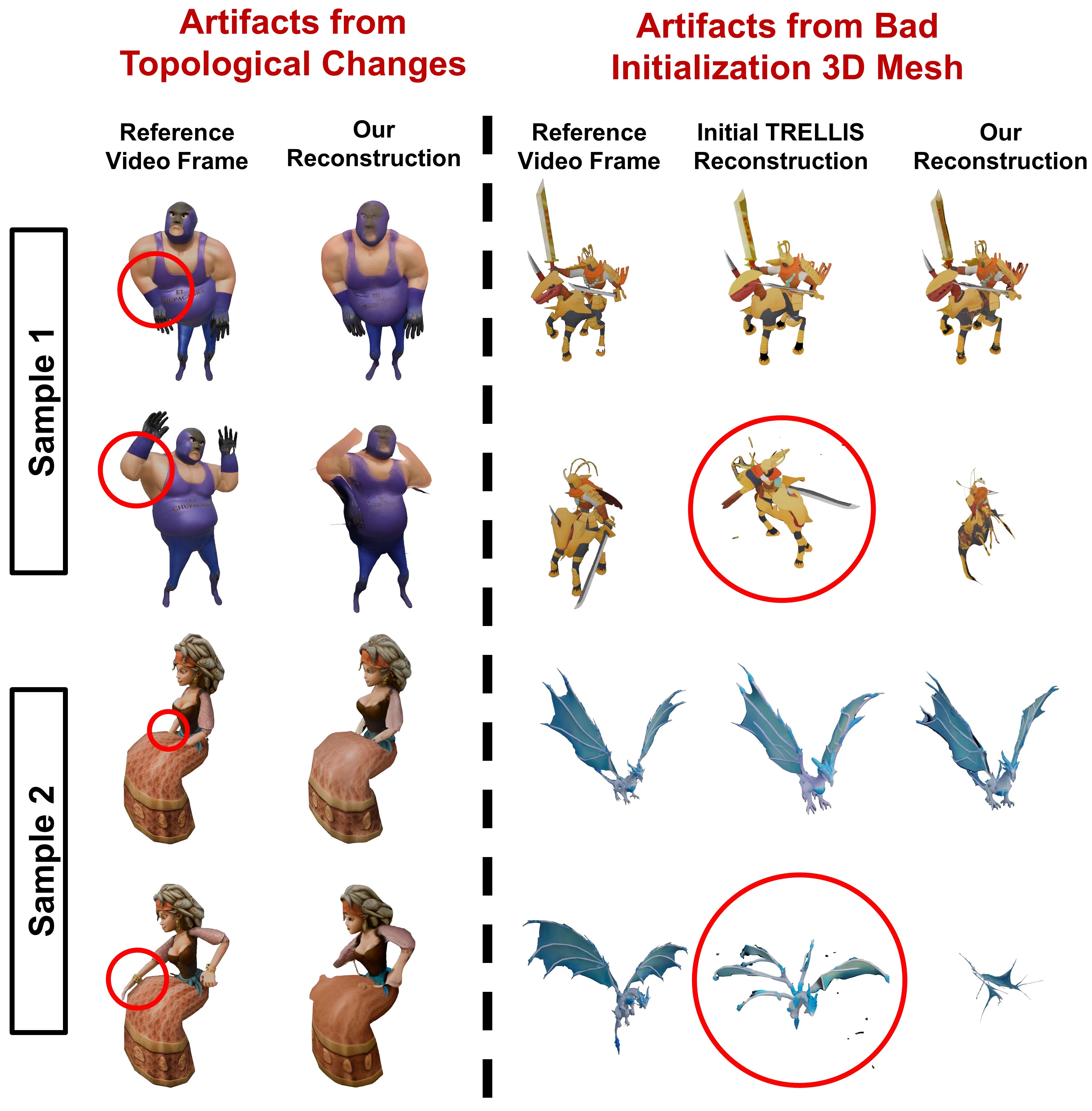}
   \caption{\textbf{Failure Cases of Our Method.} We present the limitations of our method in scenarios involving topological changes and poor 3D mesh initialization.}
   \label{fig:Limitation}
\end{figure}

\begin{figure*}[!t]
  \centering
   \includegraphics[width=\linewidth]{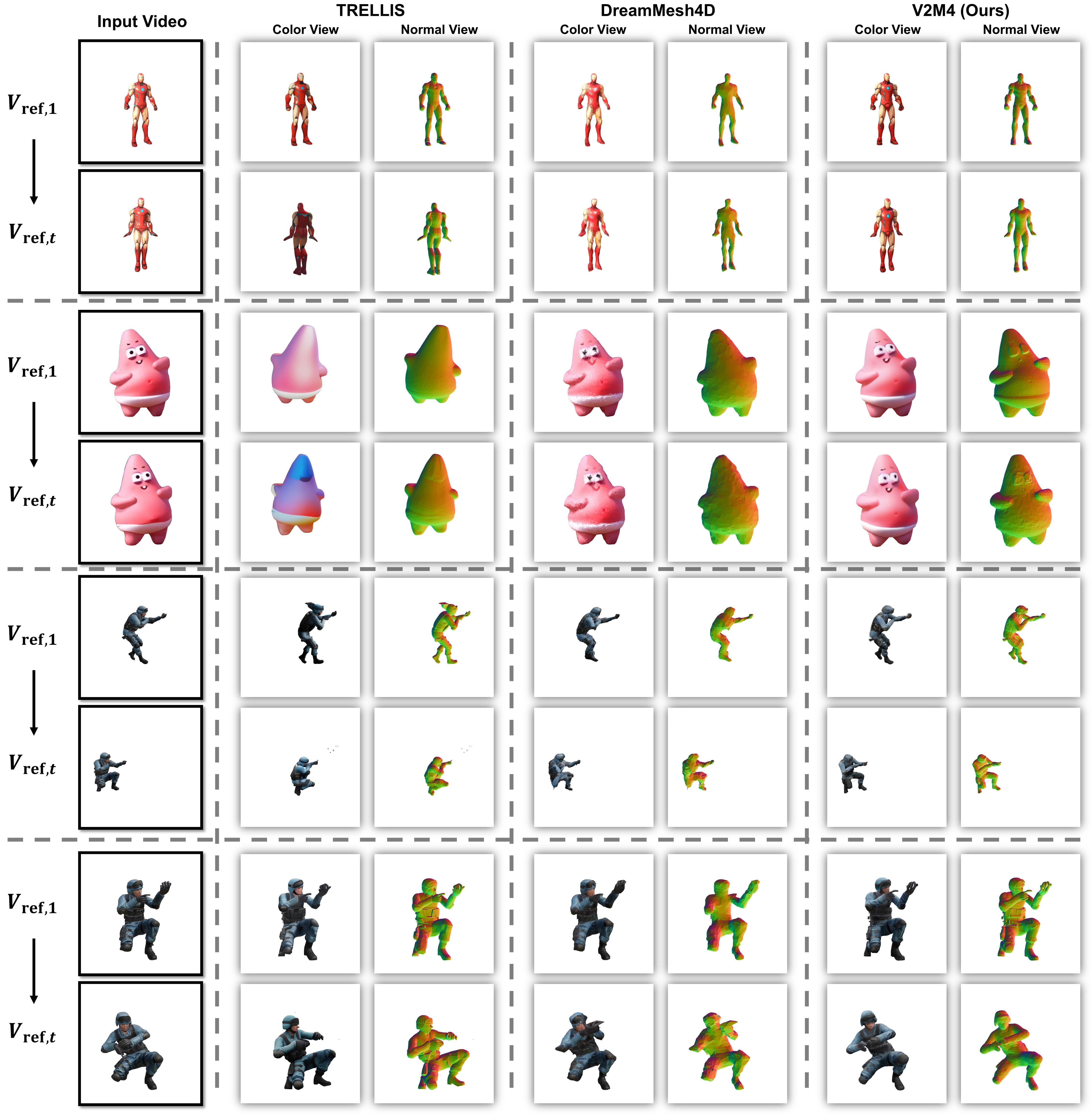}
   \caption{\textbf{More Qualitative Comparisons among Different Methods.} We present both the color view and the normal view of the reconstructed meshes. The first two samples are from the Consistent4D dataset, while the others are from our newly collected data. For each sample, two timestamps are shown for quick comparison. Please zoom in for a clearer view.}
   \label{fig:Comparison-2}
\end{figure*}

\begin{figure*}[!t]
  \centering
   \includegraphics[width=\linewidth]{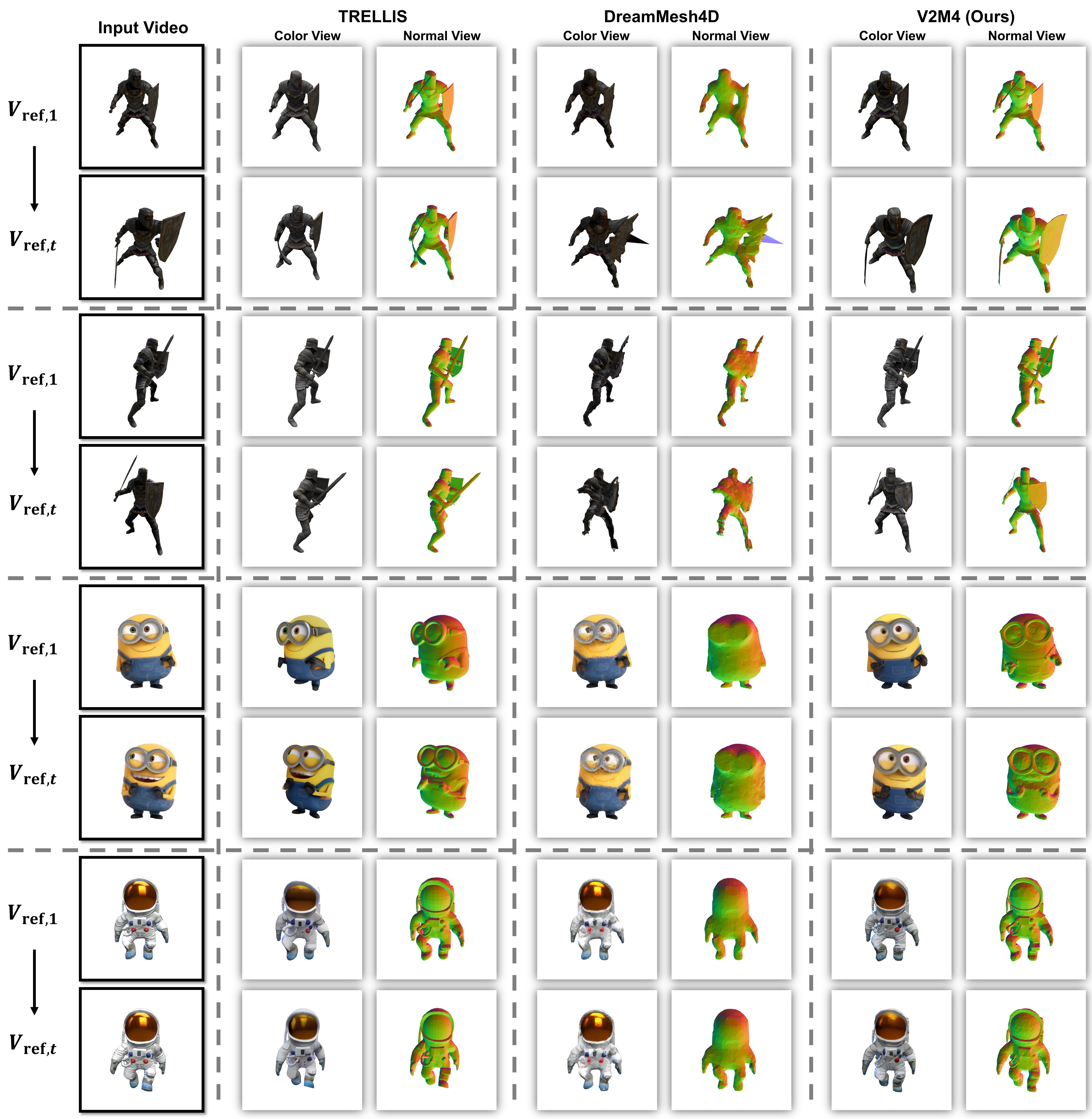}
   \caption{\textbf{More Qualitative Comparisons among Different Methods.} We present both the color view and the normal view of the reconstructed meshes. The first two samples are from the Consistent4D dataset, while the others are from our newly collected data. For each sample, two timestamps are shown for quick comparison. Please zoom in for a clearer view.}
   \label{fig:Comparison-3}
\end{figure*}

\paragraph{Mesh Appearance Refinement.} In Fig.~\ref{fig:Ablate-Appearance}, we present the mesh before and after applying our refinement strategy described in Sec.~\ref{subsec:Mesh Appearance Alignment}. It is evident that the refined mesh exhibits improved texture and is much more aligned with the appearance shown in the reference video frame.

\section{Failure Cases}
\label{append: bad cases}

In Fig.~\ref{fig:Limitation}, we display instances where our method encounters failures, including effects from poor initial 3D mesh results from TRELLIS and artifacts arising from topology changes during the animation.

\section{More Qualitative Results}
\label{append: more visual}

We display more qualitative results in Fig.~\ref{fig:Comparison-2} and Fig.~\ref{fig:Comparison-3}.

\end{document}